\documentclass[journal,10pt]{IEEEtran}

\ifx\pdfoutput\undefined
\usepackage{graphicx}  
\else
\usepackage[pdftex]{graphicx}
\fi

				\usepackage{url}
				\usepackage[numbers,sort&compress]{natbib}
				\usepackage{array}
				\usepackage{hyperref} 				
				\usepackage{stfloats}
				\usepackage{amssymb} 
				\usepackage{acronym} 
				\usepackage{booktabs}
				\usepackage{multirow} 
				\usepackage{amsmath} 
				\usepackage{amsfonts}
				\usepackage{verbatim} 
				\usepackage[utf8]{inputenc} 
				
				\usepackage{enumitem}
				\usepackage[caption=false,font=footnotesize]{subfig}
				\usepackage{amsthm} 
				\usepackage[table]{xcolor}
				\usepackage{bbding}	
				\usepackage{algorithmic} 
				\usepackage[ruled,linesnumbered]{algorithm2e}	
				\usepackage[table]{xcolor}

	\newtheorem{definition}{Definition}[section]
		\newcommand{\myvec}[1]{\ensuremath\mathbf{#1}}
		\newcommand{\spc}[1]{\ensuremath\mathbb{#1}}
		\newcommand{\set}[1]{\ensuremath\mathcal{#1}}

\begin{document}

\title{A Novel Multiobjective Cell Switch-Off\\Framework for Cellular Networks}

			\author{David~González~G.,~\IEEEmembership{Member,~IEEE,}   
							Jyri~Hämäläinen,~\IEEEmembership{Member,~IEEE,}\\
			        Halim~Yanikomeroglu,~\IEEEmembership{Senior~Member,~IEEE,}							
			        Mario~García-Lozano and Gamini Senarath  \vspace{-0.2cm}
			\IEEEcompsocitemizethanks{
			\IEEEcompsocthanksitem 
					David~González~G. and Jyri Hämäläinen are with the School of Electrical Engineering at Aalto University, Finland. Corresponding 
					email: david.gonzalez.g@ieee.org. 
					\IEEEcompsocthanksitem 
					Halim~Yanikomeroglu is with the Department of Systems and Computer Engineering, Carleton University, Ottawa, Canada. 
					\IEEEcompsocthanksitem 
					Mario~García-Lozano is with the Department of Signal Theory and Communications, Universitat Politècnica de Catalunya, Barcelona, Spain. 
					\IEEEcompsocthanksitem 
					Gamini Senarath is with Huawei Canada R\&D Centre.
					}}

\maketitle

\begin{abstract}
	Cell Switch-Off (CSO) is recognized as a promising approach to reduce the energy consumption in next-generation cellular networks. However, CSO poses serious challenges not only from the resource allocation perspective but also from the implementation point of view. Indeed, CSO represents a difficult optimization problem due to its NP-complete nature. 
			Moreover, there are a number of important practical limitations in the implementation of CSO schemes, such as the need for minimizing the real-time complexity and the number 
			of \mbox{on-off/off-on}~transitions and CSO-induced handovers.
			This article introduces a novel approach to CSO based on multiobjective optimization that makes use of the  statistical description of the service demand (known by operators). In addition, downlink and uplink coverage criteria are included and a comparative analysis between different models to characterize intercell interference is also presented to shed light on their impact on CSO. The framework distinguishes itself from other proposals in two ways: 1) The number of \mbox{on-off/off-on}~transitions as well as handovers are minimized, and 2)~the computationally-heavy part of the algorithm is executed offline, which makes its implementation feasible. The results show that the proposed scheme achieves substantial energy savings in small cell deployments where service demand is not uniformly distributed, without  compromising the Quality-of-Service (QoS) or requiring heavy real-time processing.
\end{abstract}\vspace{-0.27cm}

\IEEEpeerreviewmaketitle

\section{Introduction}\label{Sec:Intro}
\IEEEPARstart{F}{uture} hyper-dense small-cell deployments are expected to play a pivotal role in delivering high capacity and reliability by bringing the network closer to users~\cite{05:00250}.  However, in order to make hyper-dense deployments a reality, enhancements including effective interference management, self-organization, and energy efficiency are required~\cite{05:00240}. Given that large-scale deployments composed of hundreds or thousands of network elements can increase the energy consumption substantially, the need for energy efficiency (\textit{green communications}) has been recognized by the cellular communications industry as an important item in research projects and standardization activities~\cite{05:00129, 05:00183}. 

Initial attempts to improve the energy efficiency in cellular networks were oriented towards minimizing the power radiated through the air interface, which in turn reduces the electromagnetic pollution and its potential effects on human health. However, most of the energy consumption (between 50\% to 80\%) in the radio access network takes place in base stations~(BSs)~\cite{05:00183} and it is largely independent of the BSs' load. Since cellular networks are dimensioned to meet the service demand in \textit{the busy hour}~(i.e.,~peak demand), it is expected that, under non-uniform demand distributions~(both in space and time), a substantial portion of the resources may end up being underutilized, thus incurring in an unnecessary expenditure of energy. The problem may become worse in many of the scenarios foreseen for 5G, presumably characterized by hyper-dense small-cell deployments, hierarchical architectures, and highly heterogeneous service demand conditions~\cite{05:00242}. Therefore, the idea of switching off lightly loaded base stations has been considered recently as a promising method to reduce the energy consumption in cellular networks. This framework is referred to as Cell Switch-Off~(CSO) and it is focused on determining the largest set of cells that can be switched off without compromising the Quality-of-Service~(QoS) provided to users. Unfortunately, CSO is difficult to carry out due to the fact that it represents a highly challenging (combinatorial) optimization problem whose complexity grows exponentially with the number of BSs, and hence, finding optimal solutions is not possible in polynomial time. Moreover, the implementation of CSO requires coordination among neighbor cells and several other practical aspects, such as coverage provision and the need for minimizing the number of (induced) handovers and \mbox{on-off/off-on} transitions. In practice, optimizing the number of transitions, as well as the time required for them, is advisable because switching on/off BSs is far from being a simple procedure, and indeed, this process must be gradual and controlled~\cite{04:00232, 05:00126}. Moreover, a large number of transitions could result in a high number of handovers with a potentially negative impact on QoS~\cite{04:00233}. 

Although CSO is a relatively young research topic, a significant amount of contributions has been made. Hence, an exhaustive survey is both out of the scope and not feasible herein. Instead, a literature review including, in the opinion of the authors, some of the most representative works is provided. Thus, in the comparative perspective shown in Table~\ref{TableRelatedWork}, the following criteria have been considered:\vspace{-0.125cm} 
			\begin{table*}
		\caption{Summary of Related Work.}
		\vspace{-0.4cm}
		\begin{center}
		\begin{tabular}{c c c c c  c}		
		\toprule  	
			 \multirow{2}{*}{{\small \textbf{Ref} }}	&  \multirow{2}{*}{{\small \textbf{Year}}}	  &  \multirow{2}{*}{{\small \textbf{CSO type~/~architecture}}}& {\small\textbf{Realistic}} 	&  \multirow{2}{*}{{\small\textbf{Coverage aspects}}}		&  {\small\textbf{Complexity and }}	\\ 
						& 	 	& 	& {\small\textbf{ICI model}} 	& 		&  {\small\textbf{feasibility analysis}}	\\ 
		\midrule 									
									{\small \cite{7322281} }	& {\small 2016}	& {\small Snapshot  / Semi-distributed} 			&  {\small Full-load} & {\small Downlink}		&  {\small $\checkmark$}	\\			
									{\small \cite{7060678} }	& {\small 2016}	& {\small Traffic profiling  / Centralized} 			&  {\small Full-load} & {\small Downlink}		&  {\small Partially}	\\	
							{\small \cite{JaGH2016} }	& {\small 2016}	& {\small Traffic profiling  / Semi-distributed} 			&  {\small Load-coupling} & {\small Downlink}		&  {\small Partially}	\\	
					{\small \cite{7105671} }	& {\small 2015}	& {\small Snapshot / Centralized} 			&  {\small Full-load} & {\small Downlink}		&  {\small Partially}	\\	
						{\small \cite{NewHalim} }	& {\small 2015}	& {\small Snapshot / Centralized} 			&  {\small Full-load} & {\small Downlink}		&  {\small Partially}	\\	
				{\small \cite{04:00342} }	& {\small 2014}	& {\small Snapshot / Centralized} 			&  {\small Constant} & {\small Downlink}		&  {\small $\times$}	\\
				
						{\small \cite{04:00328} }	& {\small 2014}	& {\small Traffic profiling / Semi-distributed} 			&  {\small Full-load} & {\small Downlink}		&  {\small $\checkmark$}	\\
				
							{\small \cite{04:00346} }	& {\small 2014}	& {\small Snapshot / Centralized} 	&  {\small Constant} & {\small Downlink}		&  {\small $\times$}	\\
		
		{\small \cite{04:00341} }	& {\small 2013}	& {\small Snapshot / Centralized} 		&  {\small Full-load} & {\small Downlink}		&  {\small $\times$}	\\
		
		{\small \cite{04:00343} }	& {\small 2013}	& {\small Traffic profiling / Centralized} 		&  {\small Load-coupling} & {\small Downlink}		&  {\small $\times$}	\\
		
		{\small \cite{04:00344} }	& {\small 2013}	& {\small Snapshot / Centralized} 		&  {\small Constant} & {\small $\times$}		&  {\small Partially}	\\			
				{\small \cite{04:00345} }	& {\small 2013}	& {\small Snapshot / Centralized} 		&  {\small Full-load} & {\small Downlink}		&  {\small Partially}	\\
				
				{\small \cite{04:00347} }	& {\small 2013}	& {\small Snapshot / Centralized} 	&  {\small Full-load} & {\small Downlink}		&  {\small $\times$}	\\		
						{\small \cite{04:00212} }	& {\small 2012}	& {\small Snapshot / Centralized} 		&  {\small Constant} & {\small $\times$}		&  {\small $\times$}	\\		
							{\small \cite{04:00220} }	& {\small 2012}	& {\small Snapshot / Centralized} 			&  {\small Full-load} & {\small Downlink}		&  {\small $\times$}	\\						
														{\small \cite{04:00236} }	& {\small 2011}	& {\small Traffic profiling / Semi-distributed} 		&  {\small Full-load} & {\small Downlink}		&  {\small Partially}	\\							
							{\small \cite{04:00229} }	& {\small 2010}	& {\small Snapshot / Centralized} 		&  {\small Full-load} & {\small Downlink}		&  {\small $\times$}	\\
		
													{\small \cite{05:00125} }	& {\small 2010}	& {\small Snapshot / Centralized} &  {\small Constant} & {\small Downlink}		&  {\small $\times$}	\\
		
											{\small \cite{04:00225} }	& {\small 2010}	& {\small Traffic profiling / Semi-distributed} 			&  {\small Full-load} & {\small $\times$}		&  {\small $\times$}	\\		
									{\small \cite{04:00224} }	& {\small 2009}	& {\small Snapshot / Both} 	&  {\small Full-load} & {\small $\times$}		&  {\small $\times$}	\\		
			\bottomrule						 
		\end{tabular} \end{center} 
		\label{TableRelatedWork} 
		\end{table*}		
				\begin{itemize}[leftmargin=*]
					\item \textsl{CSO type / architecture}: CSO solutions can be classified as `snapshot' or `traffic profiling' CSO depending on the approach followed to take the on/off decisions. In snapshot CSO~(e.g.,~\cite{04:00342, 04:00346, 04:00341, 04:00344}),~decisions involve the analysis of discrete realizations of users, i.e., whenever a CSO decision is required, information of every single user in the network needs to be available at a central unit where an heuristic or optimization procedure is performed. Given its nature, this type of solution provides the ultimate performance in terms of energy savings. However, feasibility becomes a serious issue as it will be shown later on. On the other hand, in traffic profiling CSO~(e.g.,~\cite{04:00328, 04:00236, 04:00225}), a certain knowledge about the service demand behavior is assumed. 
				\item \textsl{Realistic ICI model}:	Switching off/on BSs modifies not only the levels of Intercell Interference~(ICI), but also the resulting load coupling~\cite{05:00169}. The load of each BS can be understood as the fraction of resources that are being used to satisfy a certain service demand. There are three different ICI models that can be used in CSO:				
				\begin{enumerate}[leftmargin=*]
					\item \textsl{Constant}: The assumption is that ICI is always constant or it does not exist~(e.g.,~\cite{04:00342, 04:00344}). In such cases, ICI levels are calculated as if all cells were always active no matter their actual state (on or off).
					\item \textsl{Full-load}~(FL): ICI is only created by active BSs, but assuming that they are fully loaded~(e.g.,~\cite{04:00345, 04:00328, 7322281}). This approach is reasonable in CSO since traffic is concentrated in a subset of BSs which tend to be highly loaded. Thus, active BSs always cause ICI to its neighbors while BSs in sleeping mode do not.
					\item \textsl{Load-coupling}~(LC): The ICI created by each BS is proportional to its load~(e.g., \cite{04:00343, JaGH2016}). This model is more realistic and accurate but involves more \mbox{complexity}. 
				\end{enumerate}
				\item \textsl{Coverage}: It indicates whether coverage aspects (in downlink, uplink, or both) are considered. This is an important criterion in CSO to avoid coverage holes. 
				\item \textsl{Complexity and feasibility}: It indicates whether a complexity and/or feasibility analysis is provided. For instance, in case of CSO, how much information is required to be exchanged among cells is an important metric. Moreover, the number of transitions and handovers in a dynamic environment must also be estimated. 
						\end{itemize}						
	As it can be seen from~Table~\ref{TableRelatedWork}, snapshot-based schemes require centralized operation, and due to the NP-complete nature of the problem, heuristics are the preferred way to deal with it. Representative examples of  this type of solutions include \cite{04:00341, 04:00212, 05:00125, 04:00342} where the main idea is, in general, to sequentially switch~1)~on highly loaded cells (cells that can get more traffic), or~2)~off lightly loaded ones (cells whose users can easily be migrated to neighbor BSs) until the service demand is fulfilled. 
	 
	In general, this type of schemes requires employing basic models for ICI, e.g., \cite{05:00125, 04:00342, 04:00212}. As indicated, another approach to CSO is traffic profiling, the case of \cite{04:00328, 04:00343, 04:00236, 04:00225}, where CSO decisions are taken assuming a certain knowledge about the service demand~\cite{04:00326, 05:00221}. However, most of the solutions presented so far employ models and assumptions oriented to macrocellular deployments, such as~\cite{04:00343, 04:00236, JaGH2016}, where the use of CSO is not so clear due to practical issues~\cite{04:00340}. Other practical aspects, such as the number of transitions and handovers are often overlooked as~well. Recently, CSO has also been studied in the context of infrastructure sharing~\cite{7105671} and by means of stochastic geometry~\cite{7060678}.
	
	Thus, in the light of these observations, this paper presents a novel multiobjective framework\footnote{A preliminary version of this work was presented in~\cite{04:00328}. From it, a US patent application has also been made: US patent application no:~14/334,134, application date: 17 July 2014.} that 1)~includes the strengths of previous proposals,~2)~overcomes many of their drawbacks, and 3)~extends the analysis to address aspects that have not fully been investigated. 
	
\vspace{0.1cm}
\begingroup
\leftskip0.99em
\rightskip0.99em
\noindent\textbf{Main contribution:} a framework for multiobjective optimization for CSO  that explicitly takes into account a statistical description of the service demand distribution when computing the performance metrics.
\par
\endgroup
\vspace{0.1cm}
As it will be illustrated by the numerical results, this idea has the following advantages:
	{\renewcommand{\labelitemi}{$\checkmark$}	
					\begin{itemize}[leftmargin=*]
	\item The use of the spatial service demand distribution (represented by a spatial probability density function) allows to the proposed algorithm to rapidly identify network topologies, i.e., on/off patterns, providing higher capacity to areas where high service demand is more likely to appear, and hence, the \textit{search space} and required computational effort is significantly reduced.
	\item Given that, in general, traffic profiles are \textit{stable} in time scales of dozens of minutes, the (computationally-heavy) optimization can be done offline and required topologies can be applied as needed as these traffic profiles are recognized/observed during network operation. This feature makes the implementation of CSO feasible, given that the required BS coordination and real-time processing are significantly reduced. However, the conceptual idea will still be valid when new paradigms (cloud computing, software defined networking, and network virtualization, see~\cite{05:00250}) allow faster computation and information exchange among network nodes. Thus, more dynamic traffic profile recognition and optimization will also be possible under the proposed framework.
	\item The proposed optimization formulation allows considering several downlink and uplink coverage criteria, such as minimum received power and Signal to Interference plus Noise Ratio~(SINR).
	\item Given that there exist a correlation between the topologies that are specific for a given traffic profile, the number of handovers and transitions is minimized.
					\end{itemize}}
Finally, the following set of secondary/minor contributions are also presented in this work:
\begin{enumerate}[leftmargin=*]
					\item Analysis of the impact of the most extended interference models (FL and LC) on the performance of CSO. 
					\item A quantitative assessment of how CSO operation affects the critical uplink power consumption (on user equipment side). To the best of the authors' knowledge, this aspect/criterion has been overlooked in most of previous studies, and only recently in \cite{NewHalim} it has been integrated within an optimization framework. 
					\item While the multiobjective problem formulation presented herein can be solved by means of standard stochastic optimization tools~\cite{08:00044}, an alternative  iterative algorithm is also proposed for computing the important tradeoff between the number of active cells and network capacity. As it will be shown shortly, although its performance is slightly inferior to stochastic search, it is significantly faster, and hence, it could be used when real-time operation becomes feasible.
\end{enumerate}
The rest of the paper is organized as follows: the next section presents the system model. The proposed Multiobjective Optimization (MO) framework (performance metrics and problem formulation) is explained in Section~\ref{Sec:PropFramework}. Section~\ref{Sec:EvaSetParameters} describes the evaluation setting and benchmarks used
in simulations. Numerical results are also analyzed therein. Section~\ref{Sec:Conclusions} closes the article with
conclusions and research directions.

\section{System Model}\label{Sec:SysModel}
	\subsection{Definitions and notations}\label{Sec:SysModel_DefNot}	
	In this study, an OFDMA cellular network is considered. The system bandwidth is $B$ and the network is composed of $L$ BSs that can be independently switched off/on. The indices of the BSs are contained in the set $\mathcal{L}=\{0,1,\cdots,L-1\}$. The set $\mathcal{A}$, composed of $A=|\mathcal{A}|$ small area elements, represents the spatial domain to which the network provides service. It is assumed that within each area elements $a\in\mathcal{A}$, the average received power is constant. The maximum transmit power per cell is $P_{\text{max}}$. The network geometry is captured by the path loss matrix $\mathbf{G}\in\mathbb{R}^{A\times L}$ (distance dependent attenuation, antenna gains, and shadowing). The vectors $\mathbf{p}_{\text{PS}}$ and $\mathbf{p}_{\text{D}}$, both $\in\mathbb{R}^{L}$, indicate the transmit power at each cell in Pilot Signals~(PS), and data channels, respectively. Cell selection is based on the average PS received power, which can be calculated by means of the following expression:
								  \begin{equation}
												\mathbf{R}_{\text{PS}} = \text{\textbf{G}} \cdot \text{diag}\left(\mathbf{p}_{\text{PS}}\odot\mathbf{x}\right),~~\mathbf{R}_{\text{PS}}\in\mathbb{R}^{A\times L}.
									\label{eq:R_CSRS}
									\end{equation}
		The operator $\odot$ denotes Hadamard (pointwise) operations. The vector $\mathbf{x}\in\{0,1\}^{L}$ indicates which cells are active and which ones are switched off. Hereafter, $\mathbf{x}$ is also referred to as 
		`network topology'. Thus, the matrix~$\mathbf{R}_{\text{PS}}$~in~(\ref{eq:R_CSRS}) contains the PS received power, i.e., $\mathbf{R}_{\text{PS}}(a,l)$ indicates the received power from the $l^{\text{th}}$ BS in the $a^{\text{th}}$ area element. Of interest is the Number of Active Cells~(NAC) in each network topology as energy consumption is related to it.				
		 The $a^{\text{th}}$ area element (the $a^{\text{th}}$ row in $\mathbf{R}_{\text{PS}}$) is served by cell~$l^{\star}$~if 
		 						\begin{equation}
										\label{Eq:ServingCell}
										l^{\star}=\underset{l\in\{0,1,\dotso,L-1\}}{\operatorname{argmax}}\hspace{0.2cm} 
										\mathbf{R}_{\text{PS}}(a,l).
								\end{equation}
		The dependence of $l^{\star}$ on $\mathbf{x}$ has not been explicitly indicated for the sake of clarity.						
		Based on (\ref{eq:R_CSRS})~and~(\ref{Eq:ServingCell}), the binary coverage matrices 
		$\text{\textbf{S}}$ and $\text{\textbf{S}}^{\text{c}}\in\mathbb{R}^{A\times L}$ can be obtained. 
		If  the $a^{\text{th}}$ pixel is served by $l^{\star}$, then $\text{\textbf{S}}(a,l^{\star})=1$. $\text{\textbf{S}}^{\text{c}}$
		is the binary complement of~$\text{\textbf{S}}$. Hence, the coverage pattern, implicitly defined in $\text{\textbf{S}}$, is a function of~$\mathbf{x}$. The cell selection rule indicated by~(\ref{Eq:ServingCell}) can be regarded as a connectivity function $f_\text{c}:\mathcal{A}\rightarrow\mathcal{L}\,\cup\,\{-1\}$. If $f_\text{c}(a)=-1$, the $a^{\text{th}}$ area element is out of coverage, i.e.,  
				\begin{itemize}[leftmargin=*]
					\item the received power in $a\in\mathcal{A}$ ($\mathbf{R}_{\text{PS}}\left(a,l^{\star} \right)$) is smaller than $P_{\text{min}}^{\text{Rx}}$, i.e., $\mathbf{R}_{\text{PS}}\left(a,l^{\star} \right)\leq P_{\text{min}}^{\text{Rx}}$,
					\item the SINR ($\psi_a$) in the area element $a\in\mathcal{A}$ is smaller than $\psi_{\text{min}}$, i.e., $\psi_a\leq \psi_{\text{min}}$, or
					\item the path loss $\mathbf{G}\left(a,l^{\star} \right)$ between the area element and its server is greater than $G^{\text{UL}}_{\text{max}}$, i.e., \mbox{$\mathbf{G}\left(a,l^{\star} \right)\geq G^{\text{UL}}_{\text{max}}$}. In practice, $G^{\text{UL}}_{\text{max}}$~is the maximum path-loss obtained from the uplink link budget (a design criterion).
				\end{itemize}	
				The cell $\mathcal{A}_l$ is the subset of $\mathcal{A}$ served by the $l^{\text{th}}$ BS. Thus, $\mathcal{A}_l\triangleq \{\,a\in\mathcal{A}:\,f_{\text{c}}(a)=l\,\}$, where $\set{A}_i\cap\set{A}_j=\emptyset,~\forall~i\neq j$. The set $\set{A}_{\text{c}}$ is the subset of $\set{A}$ that are associated to one BS. Thus, $\set{A}_{\text{c}}\triangleq \{\,a\in\set{A}:f_{\text{c}}(a)\neq-1\}=\bigcup_{l\in\set{L}} \set{A}_l$. The vector $\mathbf{\Gamma}\in\mathbb{R}^{A}$ corresponds to the spatial service demand distribution. Thus, $\mathbf{\Gamma}(a)$ indicates the probability, in the event of a new user, that the $a^{\text{th}}$ pixel has the user on it, and hence, $\mathbf{\Gamma}^{\text{T}}\cdot\mathbf{1}=1$. It should be noted that $\mathbf{\Gamma}$~is time-dependent, however it is reasonable to assume that $\mathbf{\Gamma}$ is constant during fixed intervals~\cite{04:00326}. In order to represent the service demand volume, two parameters are considered: inter-arrival time~($\lambda$) and session time~($\mu$). Both are modeled as exponentially distributed random variables. Thus, service demand's  spatial distribution and volume are described by $\mathbf{\Gamma}$ and the first order statistics of $\lambda$ and $\mu$, i.e., $\mathbb{E}\{\lambda\}$ and $\mathbb{E}\{\mu\}$, respectively. It is assumed that the QoS of a user is satisfied if the target rate ($r_{\text{min}}$) is fulfilled. Hence, the total service demand volume ($R$) in $\mathcal{A}$ is given by
 \begin{equation}
R=\sum_{a\in\set{A}}r_{a}~~~\text{[bps]},\label{Eq:V}
\end{equation}		
where 
 \begin{equation}
r_{a}=\frac{\spc{E}\{\mu\}}{\spc{E}\{\lambda\}}\cdot\mathbf{\Gamma}(a)\cdot r_{\text{min}}~~~\text{[bps]}\label{Eq:ra}
\end{equation}
corresponds to the average demand in the $a^{\text{th}}$ area element. The previous model for the service demand can easily be extended to the general case of more than one service to account with the fact that service time, inter-arrival time, spatial distribution, and target rate can be \textsl{service-specific}. Assuming that there are $N_{\text{S}}$ service classes, each of them with different characteristics, i.e., $\mu^c$, $\lambda^c$, $\mathbf{\Gamma}^c$, and $r_{\text{min}}^c$ for $c=1,2,\cdots,N_{\text{S}}$, (\ref{Eq:ra}) can be rewritten as follows:
\vspace{-0.0cm}
 \begin{equation}
r_{a}^{\text{S}}=\sum_{c=1}^{N_{\text{S}}}\left(\frac{\spc{E}\{\mu^c\}}{\spc{E}\{\lambda^c\}}\cdot\mathbf{\Gamma}^c(a)\cdot r_{\text{min}}^c\right)~~~\text{[bps]}.\label{Eq:ra_gen}
\end{equation}
The resulting spatial service demand distribution (that is required to compute the performance metrics introduced later on) can be obtained by considering the resulting demand as follows:
\vspace{-0.0cm}
  \begin{equation}
\mathbf{\Gamma}^{\text{S}}(a)=\frac{r_{a}^{\text{S}}}{\sum_{a\in\set{A}}r_{a}^{\text{S}}}.\label{Eq:Gamma_gen}
\end{equation}
Hereafter, one single service class (possibly the result of a mix of many others) is assumed for the sake of clarity, and hence, one single set of parameters ($\mu$, $\lambda$, $\mathbf{\Gamma}$, and $r_{\text{min}}$) are considered.

			\begin{definition}[Cell load]\label{Def:CellLoad}
  \normalfont The load of the $l^{\text{th}}$ BS ($\alpha_l(t)$), at any given time $t$, is defined as the fraction of the available resources (bandwidth) that are being used. 
\end{definition}		
		The average load of the $l^{\text{th}}$ BS is $\bar{\alpha}_l\triangleq\spc{E}\{\alpha_l(t)\}$. Thus, the vector~\mbox{$\bar{\boldsymbol{\alpha}}=[~\bar{\alpha}_0~\bar{\alpha}_1~\cdots~\bar{\alpha}_{L-1}~]$} indicates the load conditions in the network, on average. Note that if $\myvec{x}(l)=0$, then \mbox{$\bar{\boldsymbol{\alpha}}(l)=0$}. 	As the reader can easily infer, as long as $\bar{\boldsymbol{\alpha}}\leq\myvec{1}$, the network topology ($\myvec{x}$) is able to satisfy, on average, the service demand given by $\mathbf{\Gamma}$, $\mathbb{E}\{\lambda\}$, and $\mathbb{E}\{\mu\}$, and hence, it can be said that $\myvec{x}$ is \textit{adequate}. 
			\begin{definition}[Network capacity]\label{Def:NetCap}
			\normalfont The network capacity ($V_{\text{Cap}}$) is defined as the maximum service demand volume such that $\bar{\boldsymbol{\alpha}}\leq\myvec{1}$. Thus, $V_{\text{Cap}} \triangleq \text{max}~V:~\bar{\boldsymbol{\alpha}}\leq\myvec{1}$.
			\end{definition}			
			\begin{definition}[Saturation point]\label{Def:Sat}
			\normalfont The saturation point ($V_{\text{Sat}}$) is the minimum service demand volume such that  $\bar{\boldsymbol{\alpha}}\geq\myvec{1}$. Thus, $V_{\text{Sat}} \triangleq \text{min}~V:~\bar{\boldsymbol{\alpha}}\geq\myvec{1}$.
			\end{definition}	
		
	As indicated, different models can be used for modeling ICI. In this work, two models are considered: `\textsl{full load}' and `\textsl{load coupling}'. Recall that in full load, active cells are assumed to have full load, i.e., $\bar{\alpha}_l=1,~\text{if}~\myvec{x}(l)=1$, and $\bar{\alpha}_l=0$, if $\myvec{x}(l)=0$. In case of load coupling, the ICI created by each cell is proportional to its load. An iterative algorithm to estimate the cell load coupling  is provided in Appendix~\ref{App:LoadCoupling}. Thus, the vector $\mathbf{\Psi}\in\mathbb{R}^{A}$ representing the average SINR at each area element is given by
	\vspace{-0.55cm}
	
	{\small\begin{equation}					
					\mathbf{\Psi} = \left[ (\mathbf{S} \odot \mathbf{G}) \cdot (\mathbf{p}_{\text{D}}\odot\mathbf{x}) \hspace{0.00cm}  \right] 
						\oslash \left[\hspace{0.00cm} \left[ (\mathbf{S}^{\textnormal{c}} \odot \mathbf{G}) \cdot (\mathbf{p}_{\text{D}}\odot\mathbf{x}\odot\bar{\boldsymbol{\alpha}}) \right] \oplus 
						\sigma^2 \hspace{0.00cm}\right].
						\label{Eq:CSO_AVG_SINR}
					\end{equation}}			
The operators $\oslash$ and $\oplus$ denote Hadamard (pointwise) operations. By means of (\ref{Eq:CSO_AVG_SINR}),  average SINR figures as function of the network topology~($\myvec{x}$) are obtained. Since load levels also depend on SINR values, the load coupling generates a system of non-linear equations which have a unique non-negative ($\bar{\boldsymbol{\alpha}}\geq\myvec{0}$) solution~\cite{05:00169}. In order to estimate $\bar{\boldsymbol{\alpha}}$, let's look at the average SINR at area element level. The average SINR at $a\in\set{A}_{l}$ can be expressed as
\begin{equation}
\psi(a)=  \frac{\mathbf{p}_{\text{D}}(l)\cdot\mathbf{G}\left(a,l\right)}{\left(\displaystyle\sum\limits_{j\in\set{L}\backslash\{l\}}\bar{\alpha}_{j}\cdot\mathbf{p}_{\text{D}}(j)\cdot\mathbf{G}\left(a,j \right) \right)+\sigma^2}.  \label{Eq:SINR_Coupling_NoIntra}
\end{equation}
	In (\ref{Eq:SINR_Coupling_NoIntra}), the ICI coming from neighbor BSs is proportional to their average loads ($\bar{\alpha}_j$'s). It is customary to define link performance in terms of $\psi(a)$ by means of a concave (e.g.,  logarithmic) function ($f_{\text{LP}}$) of it, such that \mbox{$\gamma_{a}=f_{\text{LP}}(\psi(\myvec{a}))~\text{[bps/Hz]}$}. The bandwidth requirement of a single user in $a\in\set{A}_{l}$ to satisfy the QoS can be obtained as
\begin{equation}
 b_{\text{u}}(a) = \frac{r_{\text{min}}}{f_{\text{LP}}(\psi(a))}~~~\text{[Hz]},\label{Eq:UserRequirement}
\end{equation}
and the average load ($\bar{\alpha}_{l}$) in the $l^{\text{th}}$ BS would be given by
\begin{equation}
 \bar{\alpha}_{l}= \frac{1}{ B_{\text{sys}}}\cdot N_{\text{u}}^{l}\cdot b_{l}, \label{Eq:NELoadStatistically_2}
\end{equation}
where
\begin{equation}
N_{\text{u}}^{l}= \left(  \sum_{a\in\set{A}_{l}}\mathbf{\Gamma}(a) \right)\frac{\spc{E}\{\mu\}}{\spc{E}\{\lambda\}}, \label{Eq:A}
\end{equation}
and
\begin{equation}
b_{l}= \sum_{a\in\set{A}_{l}}\left(\frac{\mathbf{\Gamma}(a)}{\sum_{k\in\set{A}_{l}}\mathbf{\Gamma}(k)}\right)b_{\text{u}}(a)~~~~\text{[Hz]}. \label{Eq:B}
\end{equation}
In~(\ref{Eq:NELoadStatistically_2}), $N_{\text{u}}^{l}$ and $b_{l}$ are the average number of users and  bandwidth consumption in BS $l$, respectively. 

In order to take into account the coverage criteria and penalize solutions with coverage holes, i.e., a significant number of area elements without coverage, the spectral efficiency of the $a^{\text{th}}$ area element is stored in the vector~\mbox{$\myvec{H}\in\mathbb{R}^{A}$} and it is computed according to the following rule: \mbox{$h_a = \myvec{v}(a)\cdot f_{\text{LP}}(\psi_a)$}. The binary vector $\mathbf{v}\in\{0,1\}^{A}$ indicates if the $a^{\text{th}}$ is out of coverage. Therefore, if the $a^{\text{th}}$~area element is in outage, \mbox{$\mathbf{v}(a)$ = 1}, and 0 otherwise.

	Finally, the list of symbols is provided in Table~\ref{TableNotation}.
			\begin{table*}
		\caption{Basic notation.}
		\vspace{-0.4cm}
		\begin{center}
		\begin{tabular}{r l  r l }		
		\toprule  	
			 {\small \textbf{Symbol} }	& {\small\textbf{Description}}	& {\small \textbf{Symbol} }	& {\small\textbf{Description}} 			\\[-0.025cm]	 
		\midrule 			
			{\small $B$ }	& {\small System bandwidth}	& {\small $P_{\text{max}}$} 	& {\small Maximum transmit power per cell}			\\[-0.00cm]	
				{\small $L$ }	& {\small Number of BSs}	& {\small $\set{L}$} 	& {\small Set with the base stations' indexes}			\\[-0.00cm]	
			{\small $A$ }	& {\small Number of area elements}	& {\small $\set{A}$} 	& {\small Set of area elements in the target area}			\\[-0.00cm]	
		{\small $\myvec{G}$ }	& {\small Path-loss matrix}	& {\small $\mathbf{p}_{\text{PS}}$, $\mathbf{p}_{\text{D}}$} 	& {\small Power vectors: pilots and data channels}			\\[-0.00cm]		
				{\small $\mathbf{R}_{\text{PS}}$ }	& {\small Received power matrix}	& {\small $\mathbf{x}$} 	& {\small Network topology}			\\[-0.00cm]			
						{\small $\set{A}_{\text{c}}$ }	& {\small Coverage area ($\set{A}_{\text{c}}\subseteq\set{A}$)}	& {\small $\set{A}_l$} 	& {\small Coverage of the $l^{\text{th}}$ BS}			\\[-0.00cm]	
						{\small $\mathbf{R}_{\text{PS}}$ }	& {\small Received power matrix}	& {\small $\text{\textbf{S}}$, $\text{\textbf{S}}^{\text{c}}$} 	& {\small Coverage matrices, i.e., coverage of each cell}			\\[-0.00cm]			
							{\small $f_\text{c}$ }	& {\small Connectivity function (cell selection)}	& {\small $f_\text{LP}$} 	& {\small Link performance model}			\\[-0.00cm]	
				{\small $P_{\text{min}}$ }	& {\small Minimum received power}	& {\small $\psi_{\text{min}}$} 	& {\small Minimum SINR}			\\[-0.00cm]													
						{\small $G^{\text{UL}}_{\text{max}}$ }	& {\small Maximum path-loss}	& {\small $r_{\text{min}}$} 	& {\small Minimum target rate (QoS criterion)}			\\[-0.00cm]			
								{\small $\lambda$ }	& {\small Inter-arrival time}	& {\small $\mu$} 	& {\small Session time}			\\[-0.00cm]																			
							{\small $\mathbf{\Gamma}$ }	& {\small Spatial demand distribution}	& {\small $\bar{\boldsymbol{\alpha}}$} 	& {\small Average load vector}			\\[-0.00cm]	
							{\small $V$ }	& {\small Service demand volume}	& {\small $\mathbf{\Psi}$} 	& {\small Average SINR vector}			\\[-0.00cm]	
														{\small $\mathbf{H}$ }	& {\small Spectral efficiency vector}	& {\small $\kappa_{\text{UL}}$} 	& {\small Uplink fractional compensation}			\\[-0.00cm]		
							{\small $\mathbf{v}$ }	& {\small Coverage vector}	& {\small $\kappa_{\text{COV}}$} 	& {\small Coverage threshold}			\\[-0.00cm]				
																		{\small $\mathbf{n}$ }	& {\small Inverse of cell's size}	& {\small $$} 	& {\small }			\\[-0.05cm]		
				\bottomrule						 
		\end{tabular} \end{center} 
		\label{TableNotation} 
		\end{table*}

\section{Metrics, Problem Formulation, and Solution}\label{Sec:PropFramework}
\subsection{Multiobjective optimization: basics}\label{Sec:MObasics}	
In order to study the tradeoffs in CSO, the use of multiobjective optimization has been considered. Multiobjective optimization is the discipline that focuses on the resolution of the problems involving the simultaneous optimization of several conflicting objectives, and hence, it is a convenient tool to investigate CSO, where the two fundamental metrics, energy consumption and network capacity, are in conflict. The target is to find a subset of \textit{good} solutions $\mathcal{X}^{\star}$ from a set $\mathcal{X}$ according to a set of criteria $\set{F}=\{ f_1, f_2,\cdots,f_{|\set{F}|}\}$, with cardinality $|\set{F}|$ greater than one. In general, the objectives are in conflict, and so, improving one of them implies worsening another. Consequently, it makes no sense to talk about a single global optimum, and hence, the notion of an optimum set $\mathcal{X}^{\star}$ becomes very important. A central concept in multiobjective optimization is Pareto efficiency. A solution $\text{\textbf{x}}^{\star}\in\mathcal{X}$ has Pareto efficiency if and only if there does not exist a solution $\text{\textbf{x}}\in\mathcal{X}$, such that $\text{\textbf{x}}$ dominates $\text{\textbf{x}}^{\star}$.	A solution $\text{\textbf{x}}_{1}$ is preferred to (dominates) another solution $\text{\textbf{x}}_{2}$, ($\text{\textbf{x}}_{1}\succ\text{\textbf{x}}_{2}$), if $\text{\textbf{x}}_{1}$ is better than $\text{\textbf{x}}_{2}$ in at least one criterion and not worse than any of the remaining ones. The set $\mathcal{X}^{\star}$ of Pareto efficient solutions is called optimal nondominated set and its image is known as the Optimal Pareto Front (OPF). In multiobjective optimization, it is unusual to obtain the OPF due to problem complexity; instead, a near-optimal or estimated Pareto front (PF) is found. Readers are referred to~\cite{08:00044} for an in-depth discussion.

\subsection{Performance metrics}\label{Sec:PropFramework_ProbForm_FL}
\begin{figure}[t]
	    		\centering	    	
	    		\includegraphics[width = 0.39\textwidth]{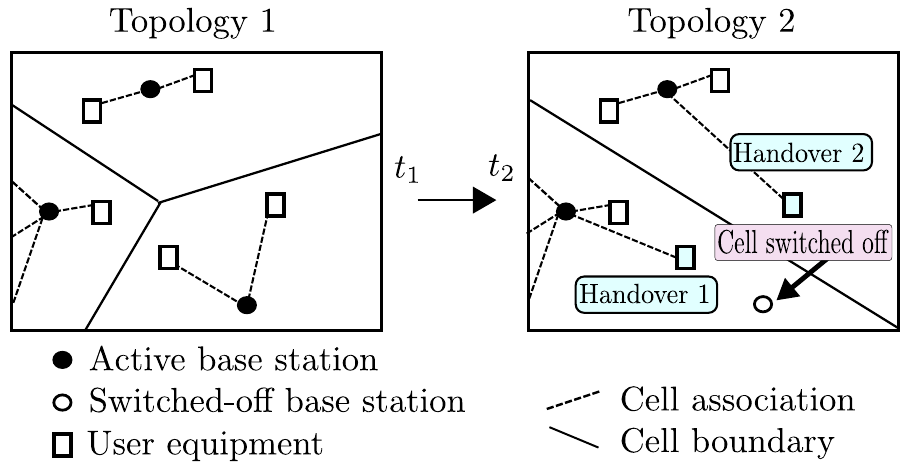}
	    		\vspace{-0.25cm}\caption{Illustration of network-initiated handover due to CSO operation.}
	    		\label{Fig:HANDOVERS}    		
			\end{figure}

 The following performance metrics have been considered\footnote{In the definition of some metrics, the dependence  with $\mathbf{x}$ is not explicit, however, it is important to note that all of them depend on $\mathbf{x}$, i.e., the network topology.}:
\begin{itemize}[leftmargin=*]
			\item \textsl{The number of active cells} ($f_1$). Under the full-load assumption, energy consumption is proportional to the number of active cells~\cite{04:00212, 05:00125}: 					
			\begin{equation}
						f_1=\mathbf{x}\cdot\mathbf{1}.
						\label{Eq:CSO_f1}
					\end{equation}							  
			\item \textsl{Average network capacity} ($f_2$).  This metric is based on the expected value of the spectral efficiency at area element level. Hence, the effect of the spatial service demand distribution ($\mathbf{\Gamma}$) must be considered. The metric is defined as follows:
			\begin{equation}
						f_2  = \left(B\cdot A\right)  \left[ \left[ ( \mathbf{H}\odot\mathbf{\Gamma} )^{\text{T}}
						\cdot\mathbf{S}\right]\odot \mathbf{n} \right]\cdot \mathbf{1}.
						\label{Eq:CSO_f2}
					\end{equation}	
														The vector \hspace{0.025cm}$\mathbf{H}\odot\mathbf{\Gamma}$ corresponds to the \textit{weighted} spectral efficiency of each area element. The idea is to give more importance to the network topologies ($\mathbf{x}$'s) that provide better aggregate capacity ($f_2$) to the areas with higher service demand. In~(\ref{Eq:CSO_f2}), $A$ (the number of area elements) is used to normalize the obtained capacity to the uniform distribution case, i.e., $\mathbf{\Gamma}(a)=1/A,\hspace{0.25cm\forall\,a\in\set{A}}$. The vector~$\mathbf{n}\in\mathbb{R}^{L}$ contains the inverse of the sum of each column in $\text{\textbf{S}}$, i.e., the number of pixels served by each cell. It is assumed that each user is served by one cell at a~time. This vector is used to distribute the capacity of each cell evenly over its coverage area, i.e., the bandwidth is shared equally by the area elements belonging to each cell. This improves the fairness in the long run similar to the proportional fairness policy that tends to share the resources equally among users as time passes. This fairness notion results in decreasing the individual rates as the number of users increases. This effect is also captured by $\mathbf{n}$ as the bandwidth per area element is inversely proportional to the size of the~cell.
				\item \textsl{Cell edge performance} ($f_3$). The $5^{\text{th}}$ percentile of the pixel rate Cumulative Distribution Function (CDF) is commonly used to provide an indicator for cell edge performance~\cite{05:00138}. A vector with the weighted average rate at area element level can be obtained as follows: 
				\begin{equation}
						\myvec{r}  = A\cdot\left( \myvec{H}\odot\myvec{\Gamma}  \right) \odot \left[ \myvec{S}\cdot \left( \myvec{n}^{\text{T}}\cdot\text{diag}(B) \right)^{\text{T}} \right].
						\label{Eq:CSO_f3}
					\end{equation}	
					Then, the percentile 5 is given by
					\begin{equation}
						f_3  = \myvec{r}^{\prime}(  0.05\cdot A).
						\label{Eq:CSO_f32}
					\end{equation}	
					The vector $\myvec{r}^{\prime}$ is a sorted (ascending order) version of $\myvec{r}$.
					\item \textsl{Uplink power consumption} ($f_4$). In order to provide an estimate of the uplink power consumption of any network topology, a fractional compensation similar to the Open Loop Power Control~(OLPC) used in Long Term Evolution~(LTE) is considered~\cite{04:00282}. It is given by
					\begin{equation}
						f_4  = \frac{1}{\sum_{k\in\set{A}_{\text{c}}}\myvec{\Gamma}(k)}\cdot\sum_{l=0}^{L-1}\sum_{a\in\set{A}_{\text{l}}}~\myvec{\Gamma}(a)\cdot\left( P_0 + \kappa_{\text{UL}} \cdot \mathbf{G}\left(a,l \right)  \right),
						\label{Eq:CSO_fUPLPC}
					\end{equation}			
								where $P_0$ is a design parameter that depends on the allocated bandwidth and target Signal-to-Noise Ratio~(SNR) and $\kappa_{\text{UL}}\in[0,1]$ is the (network controlled) fractional compensation factor. 
			\item \textsl{Load dependent power consumption} ($f_5$). In order to estimate the network power consumption under the load coupling assumption, the parameterized  BS power model proposed in~\cite{05:00232} has been used. Thus, 
			\begin{equation}
						 f_5=\sum_{l=0}^{L-1}f_{\text{PC}}^{l}(\bar{\alpha}_l),
						\label{Eq:CSO_PCLC}
					\end{equation}
						where $f_{\text{PC}}^{l}(\bar{\alpha}_l)$ is a function that gives the power consumption in the $l^{\text{th}}$~BS as function of its load. Essentially, in this model, there is a fixed power consumption ($P_0$) that is independent of the load but that can be further reduced (till $P_{\text{CSO}}$) if the base station is switched-off. Moreover, there is a part that grows linearly with the load till a maximum power consumption ($P_{\text{max}}$) that obviously contains the transmitted power over the air interface ($P_{\text{max}}^{\text{Tx}}$). 
																
			\item \textsl{Load dispersion} ($f_6$). As it will be shown, load dispersion in load coupling conditions is an important parameter because it measures how well distributed the service demand is. In order to quantify this value, the Coefficient of Variation is considered. Thus,
						\begin{equation}
	f_6= \frac{\text{std}\{\bar{\boldsymbol{\alpha}}\}}{\text{mean}\{\bar{\boldsymbol{\alpha}}\}}.
	\label{Eq:Irreg}
	\end{equation}	
		\item \textsl{Handovers}. In the context of CSO, handovers are a quite important concern~\cite{04:00233}. Handovers are produced when users need to be associated to another base station because their serving cells are switched-off. In practice, handovers are mainly produced due to users mobility, but independently of the type, either user- or network-triggered, handovers require a certain time and signaling, both at the air interface and core network. Thus, the CSO operation should, as much as possible, minimize the number of handover, i.e., the transition from one topology to another should be done with the minimal impact and/or cost. A pictorial representation of the aforementioned situation is shown in Figure~\ref{Fig:HANDOVERS}. Thus, handovers are considered herein as an important performance metric. 
		\end{itemize}
		
	\subsection{Multiobjective problem formulation}\label{Sec:PropFramework_ProbForm_MOPF}			
		The multiobjective optimization problem considered herein can be formulated as follows: 
					\begin{subequations}
					\label{OP:CSO}
					\begin{align}
					  \hspace{0.2cm} \text{optimize}~\myvec{f}(\myvec{x})& =
					         \hspace{0.025cm} [\hspace{0.05cm} 
												f_{i}\left(\mathbf{x}\right), \hspace{0.05cm} f_{j}\left(\mathbf{x}\right)  \hspace{0.025cm}], \label{OP:CSO:MAIN}\\
					    \text{subject to:} & \nonumber\\[-0.0cm]
					    \textcolor[rgb]{1,1,1}{.}&\hspace{0.5cm} \left(A^{-1}\cdot\left(\mathbf{v}^{\text{T}}\cdot\mathbf{1}\right)\right)\hspace{0.1cm}\leq\hspace{0.1cm}\kappa_{\text{COV}}, \label{OP:CSO:C1}\\[-0.05cm]
					    \textcolor[rgb]{1,1,1}{.}&\hspace{0.5cm} \mathbf{x}\in\{0,1\}^L,\hspace{0.1cm}\mathbf{x}\neq \mathbf{0}.   \label{OP:CSO:C3}
					\end{align}
					\end{subequations}	
			Problem~\eqref{OP:CSO} proposes the simultaneous optimization of two of the previously introduced performance metrics as follows:				
				\begin{itemize}
					\item Full load: if full load is assumed as model for intercell interference, $i=1$ and $j\in\{2,3,4\}$ in~\eqref{OP:CSO:MAIN}.
					\item Load coupling: if load coupling is assumed as model for intercell interference, $i=5$ and $j=6$ in~\eqref{OP:CSO:MAIN}.
				\end{itemize}
				The previous optimization scheme allows to study and characterized the tradeoffs between conflicting metrics (see Section~\ref{Sec:PropFramework_ProbForm_FL}) in a deployment-specific manner. 
 Constraints~\eqref{OP:CSO:C1}~and~\eqref{OP:CSO:C3} correspond to the coverage criterion and feasible set definition, respectively.

In general, solving multiobjective problems such as \eqref{OP:CSO} is very difficult~\cite{08:00044}. Indeed, (\ref{OP:CSO}) is a combinatorial problem that belongs to the class NP-complete, and hence, optimal solutions cannot be found in polynomial time. The domain (search space) defined by the optimization variable ($\mathbf{x}$, the on/off pattern) is a set of size $2^{L}-1$, where $L$ is the number of BSs. The objective space (or \textit{image}) is defined by the objective functions, and due to their mathematical structure, it is highly non-linear, non-convex, and full of discontinuities and local optima~\cite{08:00053}. Certain algorithms such as Simplex~\cite{05:00120} are susceptible to be trapped in local optima, while other optimization techniques, such as Sequential Quadratic Programming \cite{10:00021}, require convexity to guarantee convergence. Moreover, traditional constrained optimization, in which only one objective function is optimized subject to a set of constraints on the remaining ones, has the drawback of limiting the visibility of the whole objective space. For this reason, \textsl{heuristic}-based algorithms are popular approaches in CSO as it was seen in Section~\ref{Sec:Intro}, but unfortunately, by means of this type of solutions it is very difficult to address multiobjective optimization problems. In order to overcome this difficulty, the use of Multiobjective evolutionary algorithms~(MOEAs)~\cite{08:00052} is proposed herein as described next.

\subsection{Multiobjective evolutionary algorithms}\label{Sec:PropFramework_ProbForm_MOEAs}	
As it was mentioned, heuristic solutions are usually problem-specific and typically used for single-objective optimization. Thus, the so-called `\textsl{metaheuristics}' have become an active research field~\cite{08:00053}. Metaheuristics can be used to solve very general kind of multiobjective optimization problems, such as the CSO formulation presented herein. 
Indeed, \eqref{OP:CSO} requires a tool able to~1)~find good solutions by efficiently exploring the search space, and~2)~operate efficiently with multiple criteria and a large number of design variables. In addition, it should not have strong requirements, such as convexity or continuity. Multiobjective evolutionary algorithms~(MOEAs)~\cite{08:00052} fulfill the previous goals, and hence, their use is proposed to deal with the CSO framework presented herein. MOEAs are population-based metaheuristics that simulate the process of natural evolution and they are convenient due to their \textit{black-box} nature that requires no assumption on the objective functions. 

Thus, the Nondominated Sorting Genetic Algorithm II~(NSGA-II)~\cite{05:00087} is employed herein to solve~(\ref{OP:CSO}). NSGA-II is accepted and well-recognized as a reference in the field of evolutionary optimization as it has desirable features, such as \textit{elitism} (the ability to preserve good solutions), and mechanisms to flexibly improve \textit{convergence} and \textit{distribution}. Further details can be found in~\cite{08:00052}. One key insight for selecting evolutionary (genetic) algorithms is that, in CSO, a certain correlation is expected among network topologies that are suitable for a given spatial service demand distribution, i.e., they are expected to be similar, with more cells where the traffic is concentrated. The operation in evolutionary algorithms precisely does that, i.e., once a good (Pareto efficient) solution (network topology) is found, the algorithm iteratively try to improve it by~1)~combining it with other good solutions (\textsl{crossover} mechanism), and 2)~adding random minor variations to them (\textsl{mutation} mechanism). The complete description of NSGA-II can be found in~\cite{05:00087}. 
As it will be shown, the use of MOEAs provides a quite convenient approach to CSO. However, depending on the scale of the problem, convergence can be slow, especially if computational resources are limited. Thus, based on the insight previously indicated, and in order to provide additional possibilities, Algorithm~\ref{Alg:CSO:MinDist} is also proposed for solving (finding the set $\mathcal{X}^{\star}$ of Pareto efficient solutions) a particular, yet important, case of (\ref{OP:CSO}); when the number of active cells ($f_1$) and the average network capacity ($f_2$) need to be jointly optimized. Given that the need for minimizing the number of transitions is very important from a practical point of view, Algorithm~\ref{Alg:CSO:MinDist} aims at finding a collection of network topologies, all with different number of active BSs, featuring~1)~the minimum distance property, and~2)~acceptable performance. In this context, the word~\textit{distance} refers to the Hamming distance ($d_{\text{H}}$), i.e., the number of positions in which the corresponding symbols in two different solutions are different. In this manner, for two solutions $\mathbf{x}_i$ and $\mathbf{x}_j$ in a set $\mathcal{X}_{\text{MD}}^{\star}$ featuring the minimum distance property, \mbox{$d_{\text{H}}\left(\mathbf{x}_i,\mathbf{x}_j\right)=1\Rightarrow|(\mathbf{x}_i\cdot\mathbf{1})-(\mathbf{x}_j\cdot\mathbf{1})|=1$} always holds. Initially, Algorithm~\ref{Alg:CSO:MinDist} determines the best~topology with~1~active BS~($\mathbf{x}_1$) in line~2. Then, in lines~4-14, for each successive number of active cells ($\text{NAC}=2,~\dotso,~L$), the algorithm sequentially finds the BSs that should be activated (resulting in the solution $\mathbf{x}_j$), such that~1)~the Hamming distance with the previous solution $\mathbf{x}_{j-1}$ is one, and 2)~the function $f_2$ is maximized. Thus, each solution added to $\mathcal{X}^{\star}_{\text{MD}}$ provides the biggest increment in terms of $f_2$ with respect to the one previously added, and only one off/on transition is required. It should be noted that, although not explicitly indicated, Algorithm~\ref{Alg:CSO:MinDist} indeed optimizes not only the number of active base stations and the network capacity, but also the number of transitions when moving from one topology to another. Thus, more than two objetives are jointly considered. The same applies for~\eqref{OP:CSO}, i.e., more than two metrics could be considered, at expense of an increase in complexity, but it should be taken into account that in the context of CSO, a Pareto Front in more than two dimensions could complicate the implementation.    
							\begin{algorithm}[t]
								\caption{Minimum Distance Algorithm~(MDA).}
								\label{Alg:CSO:MinDist}
								\SetAlgoLined							
								\SetKwInOut{Input}{input}
								\SetKwInOut{Output}{output}	
								\SetKwFunction{BestBS}{BestBS}
								\BlankLine\small
							\Input{{\small
											$\mathcal{X}_1$: $\mathcal{X}_1=\{ \mathbf{x}\in\mathcal{X}
											\hspace{0.075cm}|\hspace{0.075cm}\mathbf{x}\cdot\mathbf{1}=1\}$, $|\mathcal{X}_1|=L$.}} \vspace{-0.0cm}
							\Output{{\small $\mathcal{X}^{\star}_{\text{MDA}}$: A set of $L$ network topologies.	}}
							{\small
							$C^{\star}\leftarrow0$;~~$\mathcal{X}^{\star}_{\text{MD}}\leftarrow \emptyset$\;\vspace{-0.0cm}
							$\mathbf{x}_1\leftarrow$\BestBS{$\mathcal{X}_1$}\;\vspace{-0.0cm}
							$\mathcal{X}^{\star}_{\text{MD}}\leftarrow \mathcal{X}^{\star}_{\text{MD}} \cup\{\mathbf{x}_1\}$\;\vspace{-0.0cm}
							\For{\textnormal{\textbf{each}} $j=2:L$}
							{\vspace{-0.0cm}
									$C^{\star}\leftarrow0$\;\vspace{-0.0cm}
									\For{\textnormal{\textbf{each}} $\mathbf{x}\in\mathcal{X}_{j}\hspace{0.1cm}|\hspace{0.1cm}d_{\text{H}}(\mathbf{x},\mathbf{x}_{j-1})=1$}
									{\vspace{-0.0cm}
											$C_\mathbf{x}\leftarrow f_2(\mathbf{x})$\;\vspace{-0.0cm}
											\If{$C_\mathbf{x} > C^{\star}$}
											{\vspace{-0.0cm}
													$C^{\star}\leftarrow C_\mathbf{x}$\;\vspace{-0.0cm}
													$\mathbf{x}_j\leftarrow \mathbf{x}$\;\vspace{-0.0cm}
											}		\vspace{-0.0cm}									
									}	\vspace{-0.0cm}
									$\mathcal{X}^{\star}_{\text{MD}}\leftarrow \mathcal{X}^{\star}_{\text{MD}} \cup\{\mathbf{x}_j\}$\;\vspace{-0.0cm}
									
							}\vspace{-0.0cm}
							
							\textbf{return} $\mathcal{X}^{\star}$\;\vspace{-0.0cm}
							}
							\end{algorithm}	

\subsection{Conceptual design and implementation}\label{Sec:PropFramework_ConceptualDesign}
\begin{figure}[t]
	    		\centering	    	
	    		\includegraphics[width = 0.48\textwidth]{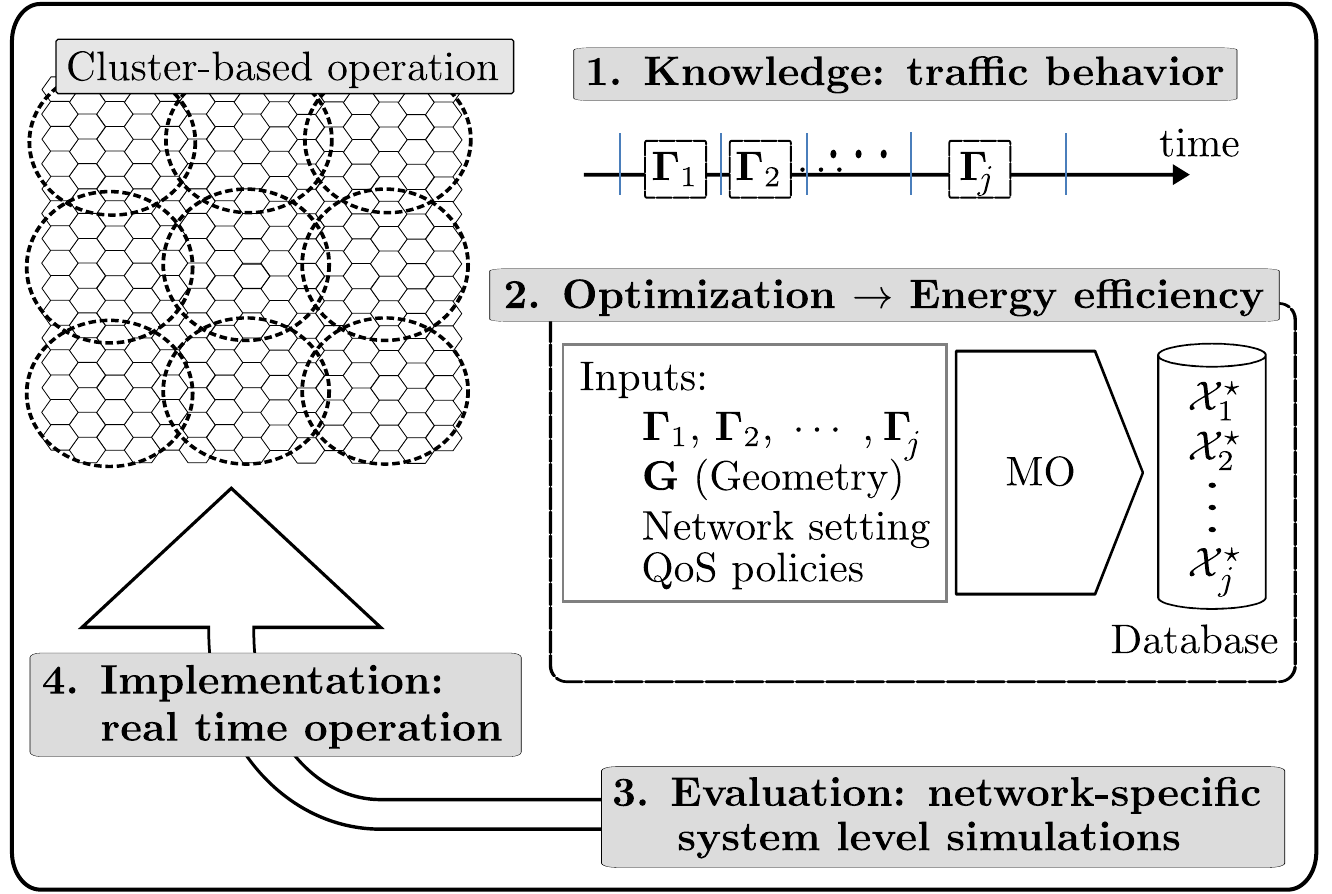}
	    		\vspace{-0.25cm}\caption{Conceptual design of the MO framework for CSO.}
	    		\label{Fig:CONCEPTUAL_DESIGN}    		
			\end{figure}

				 Figure~\ref{Fig:CONCEPTUAL_DESIGN} illustrates the conceptual design of the proposed multiobjective framework. The framework relies on having a statistical description of the behavior of the service demand (in time and space). Thus, by means of different traffic distributions ($\mathbf{\Gamma}_x$), the spatial component of the traffic at different moments of the day can be captured. These patterns can be considered fairly constant during time intervals
					of \textit{small} duration (tens of minutes or few hours)~\cite{04:00225, 04:00236}.
					Starting from the knowledge of a given $\mathbf{\Gamma}$, network analysis and optimization based on (\ref{OP:CSO}) is done offline. The main idea is that, for different demand conditions (spatial distribution and volume), different sets of Pareto efficient network topologies can be obtained, i.e., for each $\mathbf{\Gamma}_x$, there is a corresponding $\mathcal{X}^{\star}_x$. These sets of near-optimal solutions ($\mathcal{X}^{\star}_x$'s) can be 
					evaluated by means of system level simulations (in which several QoS criteria, scheduling policies, and ICI models can be considered independently) in order to determine which network topologies ($\myvec{x}\in\mathcal{X}^{\star}_x$) provide the desired level of QoS. Obviously, the network operator may act rather conservatively in this selection process as it will be explained in Section~\ref{Sec:EvaSetParameters}.	
					Moreover, in order to allow for semi-distributed implementation, a cluster-based operation is encouraged. The benefit of doing so is twofold. First, the demand in relatively small areas covered by small cells (e.g., pico-cells in a university campus) can be characterized easily. Second, the amount of intercell coordination is reduced compared with the schemes aiming at operating in large urban areas. Since demand profiles are stored and indexed at coordinating points in each cluster, the amount of data that need to be exchanged (from time to time) is negligible. Instead, different clusters (a certain amount of overlapping can be allowed) can also share information in longer time scales,  so that better decisions can be made in boundary cells. In any case, the idea of identifying traffic profiles and applying multiobjective-optimized on/off patterns, is compatible with novel paradigms that 1)~are being considered for 5G (\textit{cloud-networking} and \textit{virtualization}~\cite{05:00242, 05:00240}) and~2)~would allow for more dynamic and centralized operation. In addition, several research contributions in the increasingly research field of service demand modeling and pattern recognition~\cite{6966182, 7147784, 6757900} (a research problem out of our scope) are appearing, and hence, the method proposed herein can extensively be benefit from that activity.

\section{Performance Evaluation}\label{Sec:EvaSetParameters}
 	\subsection{Simulation conditions and parameters}\label{Sec:EvaSetParameters_SimulPar} 	

									\begin{figure}[t]
	    		\centering	    	
						\subfloat[Cellular layout.]
	    		{\label{Fig:LAYOUT}
	    		\includegraphics[width = 0.23\textwidth]{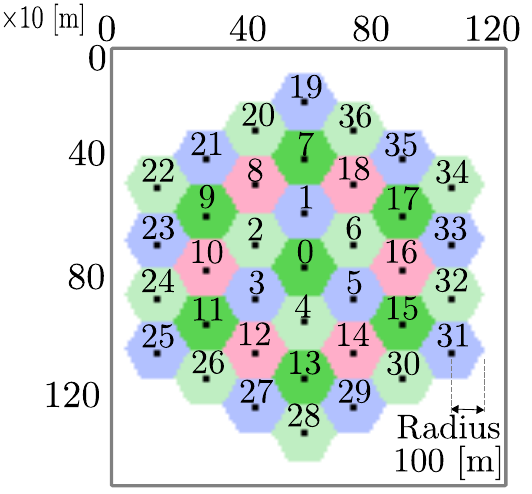}}			
							\subfloat[Spatial service demand ($\mathbf{\Gamma}$).]
	    		{\label{Fig:STD}
	    		\includegraphics[width = 0.23\textwidth]{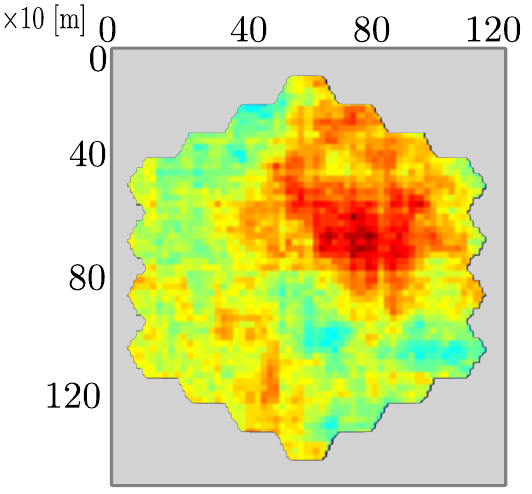}}\\[-0.2cm]	
					\subfloat[CDF of pixel prob. ($\mathbf{\Gamma}(a)$).]
	    		{\label{Fig:STD_CDF}
	    		\includegraphics[width = 0.23\textwidth]{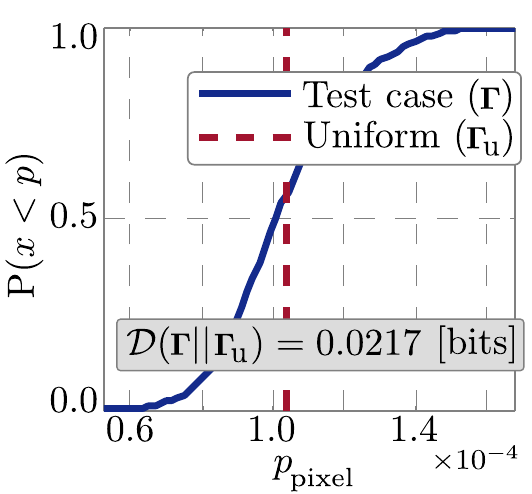}}	
					\subfloat[Power consumption model.]
	    		{\label{Fig:POWER_CONSUMPTION_MODEL}
	    		\includegraphics[width = 0.224\textwidth]{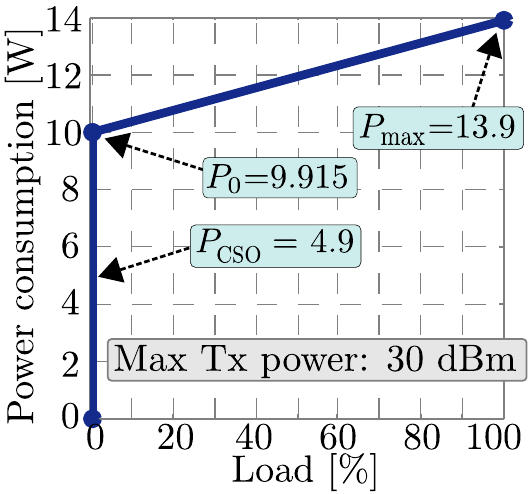}}
	    		\vspace{-0.05cm}\caption{Test case scenario.}
	    		\label{Fig:TESTCASE}    		
			\end{figure}

				The simulation setup is based on the assumptions for evaluating the IMT-Advanced systems~\cite{03:00002}. The urban micro-cell (UMi) downlink scenario was chosen. Fig.~\ref{Fig:LAYOUT} shows the corresponding cellular layout. As it can be seen, the network is composed of 37 small cells (radius = $100\,$m, $\text{network area}\approx1\,\text{km}^2$). Fig.~\ref{Fig:STD} corresponds to the (irregular) spatial service demand distribution ($\myvec{\Gamma}$) used in the numerical examples. The Kullback-Leiber distance $\set{D}$ with respect to the uniform distribution~($\myvec{\Gamma}_{\text{u}}$) can be used as a measure of the non-uniformity of the spatial service demand distribution as it is shown in Figure~\ref{Fig:STD_CDF}. This setting is perfectly valid to study CSO as in this context gains are obtained from the mismatch between demand and supply. Indeed, CSO is about finding the smallest network topology that is \textit{compatible}~\cite{7127727} enough with the service demand to provide the required QoS and coverage.
				
				In this study, a Single-Input Single-Output~(SISO) is considered. This assumption does not imply any loss of generality as long as all cells use the same scheme. The load-dependent power consumption model, based on the parameters given in~\cite{05:00232}, for pico-BSs assuming an operating bandwidth of $5\,$MHz and a maximum transmission power of $30\,$dBm is shown in~Fig.~\ref{Fig:POWER_CONSUMPTION_MODEL}.		
					
			Dynamic system level simulations are carried based on Monte~Carlo experiments. The results compile statistics taken from 100~independent experiments each of which has a duration of $5400\,$s. At each cell, the scheduler assigns each user with a bandwidth such that the target rate ($r_{\text{min}}$) is satisfied. If the percentage of users that obtain a rate equal to $r_{\text{min}}$ is greater or equal to the operator-specific target QoS ($Q$), then the QoS policy is said to be fulfilled. Thus, in order to satisfy the maximum number of users, users are sorted based on their spectral  efficiency and served accordingly. When there is not enough bandwidth to satisfy a user, the resource allocation ends. The set of parameters used in simulations is provided in Table~\ref{TableEvaluationSetting}. Calibration and complexity aspects of NSGA-II are briefly discussed in Section~\ref{Sec:NumResults_Complexity analysis}, and additional guidelines can be found in~\cite{05:00087, 05:00138}. The experimentally obtained setting is also shown in Table~\ref{TableEvaluationSetting}.

 														\begin{table*}
		\caption{Evaluation setting and parameters.}
		\vspace{-0.4cm}
		\begin{center}
		\begin{tabular}{r l |r l| r l}		
		\toprule  
			 \multicolumn{6}{c}{{\scriptsize  \textbf{General setting}}} 		\\ 		
			{\scriptsize  Number of cells~($L$)}	& {\scriptsize 37~(wraparound, omni)}	& {\scriptsize Carrier freq.} 	& {\scriptsize 2.140~GHz} & {\scriptsize Bandwidth~($B$)} 	& {\scriptsize 5~MHz}			\\[-0.01cm]	
			{\scriptsize  Max. BS transmit power~($P_{\text{max}}^{\text{Tx}}$)}	& {\scriptsize 30~dBm}	& {\scriptsize Pixels' resolution} 	& {\scriptsize $5\times 5$~m$^2$} & {\scriptsize Path loss} 	& {\scriptsize M.2135 Umi~\cite{03:00002}}			\\[-0.01cm]		
			{\scriptsize  BS's height}	& {\scriptsize 15~m}	& {\scriptsize Noise power~($\sigma^2$)} 	& {\scriptsize -174~dBm/Hz} & {\scriptsize Shadowing } 	& {\scriptsize $\set{N}(0.4)~[\text{dB}]$ }			\\[-0.01cm]		
			{\scriptsize  Cell selection ($f_c$)}	& {\scriptsize Highest Rx. power}	& {\scriptsize Rx. power~($P_{\text{min}}$)} 	& {\scriptsize -123 dBm} & {\scriptsize  Small scale fad.} 	& {\scriptsize As in~\cite{03:00002}}			\\[-0.01cm]		
			{\scriptsize  Link performance ($f_{\text{LP}}$)}	& {\scriptsize Shannon's formula}	& {\scriptsize Frac. comp. ($\kappa_{\text{UL}}$)} 	& {\scriptsize 1.00} & {\scriptsize Cov. ($\kappa_{\text{COV}}$)} 	& {\scriptsize 0.02}			\\[-0.01cm]		
			{\scriptsize  Max. path loss ($G_{\text{max}}^{\text{UL}}$)}	& {\scriptsize 163 dB}	& {\scriptsize Min. rate~($r_{\text{min}}$)} 	& {\scriptsize 400 kbps} & {\scriptsize Target QoS ($Q$)} 	& {\scriptsize 97.5$\%$}			\\[-0.01cm]		
			{\scriptsize  User distribution}	& {\scriptsize According to $\myvec{\Gamma}$}	& {\scriptsize Traffic model} 	& {\scriptsize Full buffers} & {\scriptsize SINR~($\psi_{\text{min}}$)} 	& {\scriptsize -7.0 dB}			\\[-0.01cm]	
			{\scriptsize  QoS checking interval}	& {\scriptsize $1\,\text{s}$}	& {\scriptsize } 	& {\scriptsize } & {\scriptsize } 	& {\scriptsize }			\\[-0.01cm]				
						\multicolumn{6}{c}{{\scriptsize  \textbf{Calibration of NSGA-II}}}\\	
{\scriptsize  Population size}	& {\scriptsize 100}	& {\scriptsize Crossover. prob.} 	& {\scriptsize 1.00} & {\scriptsize Type of var.} 	& {\scriptsize Discrete}			\\[-0.01cm]																				{\scriptsize  Mutation prob.}	& {\scriptsize $1/L$}	& {\scriptsize Termination crit.} 	& \multicolumn{3}{l}{{\scriptsize Hypervolume $<0.001\%$, \cite{08:00044}}}		\\[-0.01cm]							
				\bottomrule						 
		\end{tabular} \end{center} 
		\label{TableEvaluationSetting} 
		\end{table*}	

				\vspace{-0.0cm}\begin{figure}
	    		\centering
	    		\subfloat[Coverage maps.]
	    		{\label{Fig:CoverageMaps}
	    		\includegraphics[width = 0.235\textwidth]{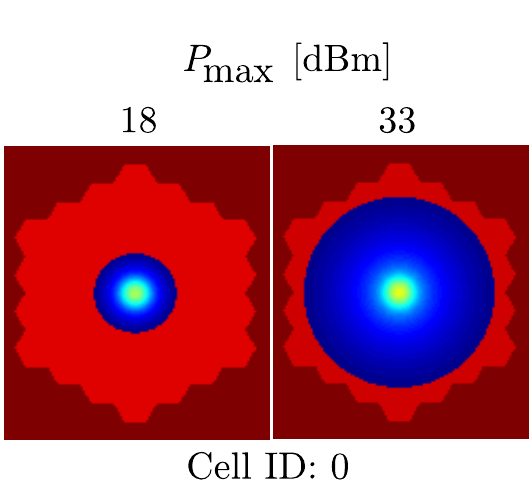}}
					\subfloat[Coverage vs. Tx power.]
	    		{\label{Fig:CoverageVsTxPower}
	    		\includegraphics[width = 0.235\textwidth]{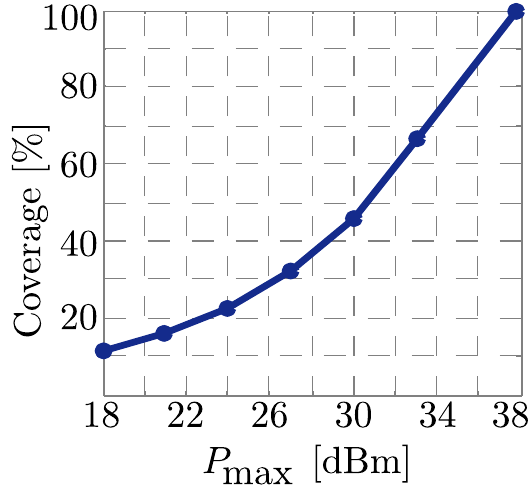}}\\[-0.2cm]					
					\subfloat[Detectable cells.]
	    		{\label{Fig:ServersOverlapping2}
	    		\includegraphics[width = 0.235\textwidth]{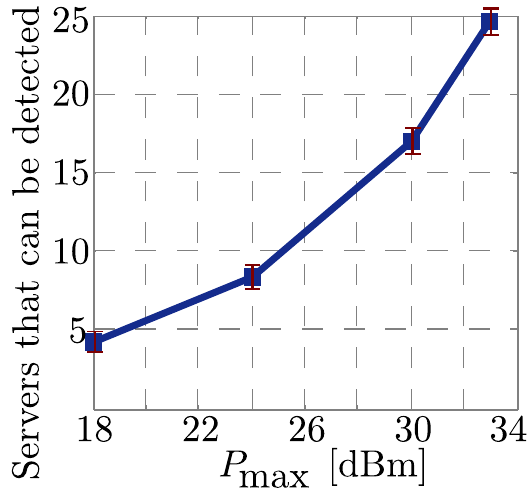}}
					\subfloat[Candidate servers.]
	    		{\label{Fig:ServersOverlapping11}
	    		\includegraphics[width = 0.235\textwidth]{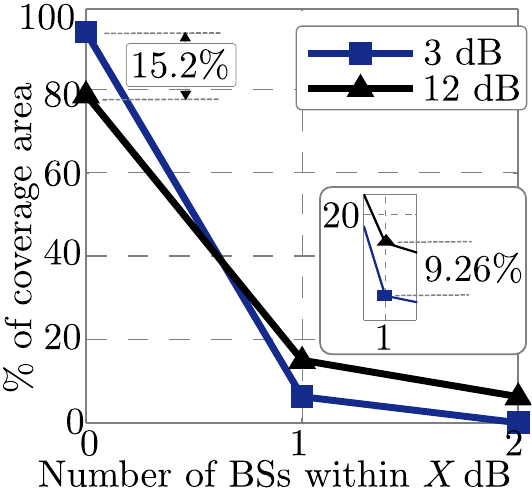}}
	    		\vspace{-0.1cm}\caption{Coverage aspects: impact of transmit power.}
	    		\label{Fig:CoverageTxPow}    		
			\end{figure}	

	\subsection{Coverage aspects}\label{Sec:NumResults_Cov}
The first part of this section is devoted to illustrate some coverage aspects and provides insights into the potential impact of the transmit power on the performance of CSO. Fig.~\ref{Fig:CoverageMaps} provides a qualitative perspective. The figure shows the size of the maximum coverage (points in which the received PS power is greater than $P_{\text{min}}$) for the central BSs ($l=0$) for two different transmit powers ($P_{\text{max}}=18\,$dBm and $P_{\text{max}}=33\,$dBm). For the sake of clarity, shadowing is not considered. A quantitative description is shown in Fig.~\ref{Fig:CoverageVsTxPower} which indicates the percentage of the target area~($\set{A}$) that can be covered with different values of $P_{\text{max}}$. Note for instance that, starting from $18\,$dBm~($15\,\%$ of coverage), $P_{\text{max}}$ need to be increased more than eight times (up to $30\,$dBm) to double the coverage (up to $30\,\%$), while reaching $60\,\%$ of coverage requires less than four times the power required for $30\,\%$ of coverage. Obviously, this depends on the propagation model, but the message is that this analysis should be taken into account during the design phase of any CSO strategy in order to determine appropriate values for $P_{\text{max}}$. In the results shown in Figs.~\ref{Fig:ServersOverlapping2}~and~\ref{Fig:ServersOverlapping11}, all the cells are active and transmit at the same~$P_{\text{max}}$. Fig.~\ref{Fig:ServersOverlapping2} indicates the average number of BS that can be \textit{detected} as a function of $P_{\text{max}}$~(the average is taken over the whole coverage area). Fig.~\ref{Fig:ServersOverlapping11} shows the percentage of the coverage area in which $x$ BSs (servers) are \textit{heard} with a quality~(SINR) within $X\,$dB below the one of the best server. From these results, it becomes clear that the choice of $P_{\text{max}}$ has a big influence on the size of the feasible set in (\ref{OP:CSO}), i.e., the set of $\myvec{x}$'s for which Constraint~\ref{OP:CSO:C1} is fulfilled. Hence, the impact of $P_{\text{max}}$ is significant, mainly in low load conditions.

\subsection{Estimation of network topologies}\label{Sec:NumResults_NetTop}
				\begin{figure}
	    		\centering
	    		\subfloat[Capacity: MOEA vs. MDA.]
	    		{\label{Fig:PARETO_FRONT_MOEA_MDA}
	    		\includegraphics[width = 0.235\textwidth]{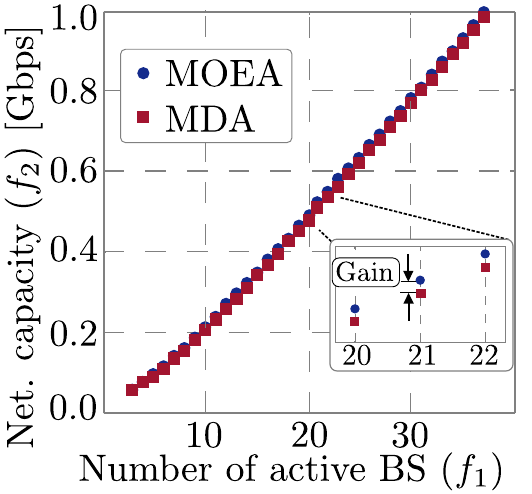}}
					\subfloat[Gains: MOEA vs. MDA.]
	    		{\label{Fig:PARETO_FRONT_MOEA_MDA_GAINS}
	    		\includegraphics[width = 0.235\textwidth]{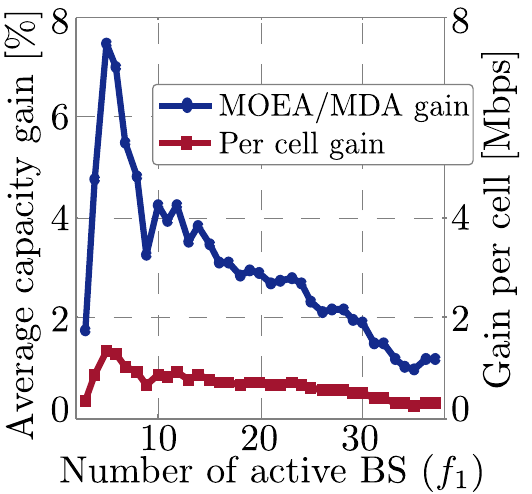}}\\[-0.2cm]	
					\subfloat[Cell edge performance.]
	    		{\label{Fig:PARETO_FRONT_P5_NEL}
	    		\includegraphics[width = 0.235\textwidth]{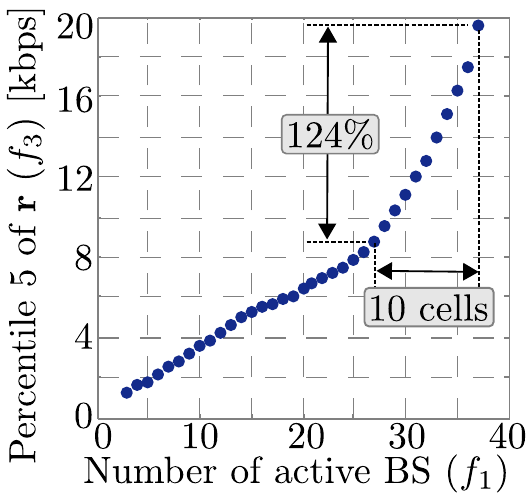}}
					\subfloat[Uplink Tx power.]
	    		{\label{Fig:PARETO_FRONT_UPC_NEL}
	    		\includegraphics[width = 0.235\textwidth]{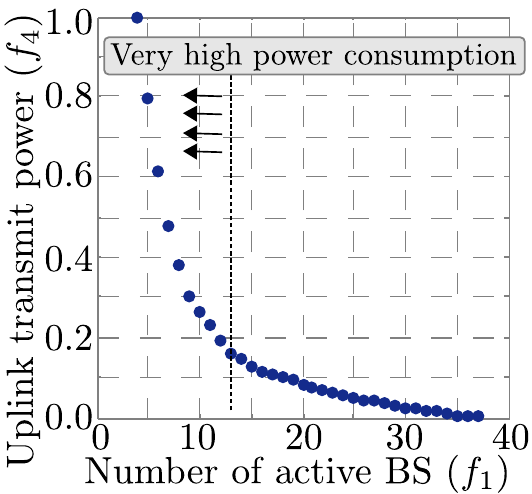}}
	    		\vspace{-0.03cm}\caption{Multiobjective optimization results.}
	    		\label{Fig:NO_CELL_COUPLING_PFs}    		
			\end{figure}	
						\begin{figure*}[t]
	    		\centering
					\subfloat[Load sharing.]
	    		{\label{Fig:CV_LOADS}
	    		\includegraphics[width = 0.26\textwidth]{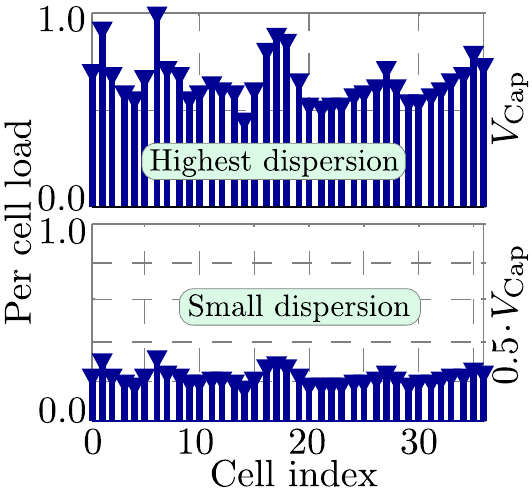}}
	    		\subfloat[Impact of load.]
	    		{\label{Fig:CV_PC_LOAD}
	    		\includegraphics[width = 0.26\textwidth]{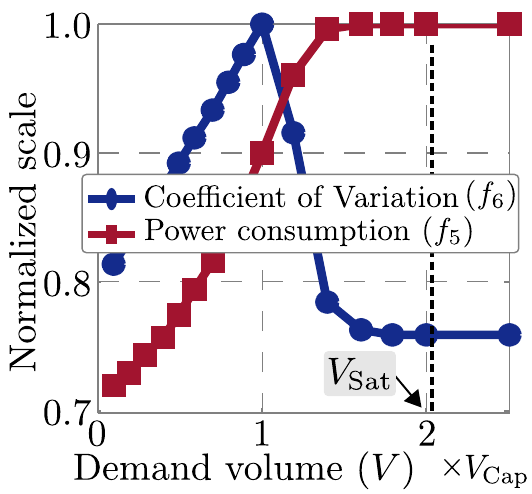}}					
					\subfloat[MO (Load = $0.6\cdot C_{\text{Max}}$).]
	    		{\label{Fig:CV_PC_MO}
	    		\includegraphics[width = 0.26\textwidth]{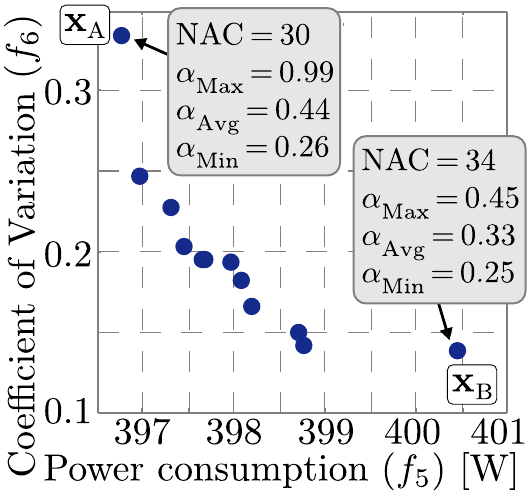}}
	    		\vspace{-0.10cm}\caption{Analysis considering cell coupling and load-dependent power consumption.}
	    		\label{Fig:CELL_COUPLING_BASED_ANALYSIS}    		
			\end{figure*}		
First, the results regarding the solution of~(\ref{OP:CSO}) for the objectives functions introduced in Subsection~\ref{Sec:PropFramework_ProbForm_FL}~($f_1,\,f_2,\,f_3,\,\text{and}\,f_4$) are provided. Fig.~\ref{Fig:PARETO_FRONT_MOEA_MDA} shows the resulting Pareto Front by solving~(\ref{OP:CSO}), when $i=1\,\text{and}\,j=2$ in~(\ref{OP:CSO:MAIN}), i.e., the joint optimization of the number of active BS ($f_1$) and the average network capacity~($f_2$), by means of MOEAs (algorithm NSGA-II) and Algorithm~\ref{Alg:CSO:MinDist}. As expected, the use of evolutionary optimization provides \textit{better} solutions than Algorithm~\ref{Alg:CSO:MinDist}, i.e., greater values of $f_2$ for the same value of $f_1$. However, it is important to recall that the solutions obtained through Algorithm~\ref{Alg:CSO:MinDist} feature the minimum distance property (see Section~\ref{Sec:PropFramework_ProbForm_MOPF}), and that, Algorithm~\ref{Alg:CSO:MinDist}~($\mathcal{O}(L^2)$) is, in case of small-to-moderate cluster size, less complex than NSGA-II ($\set{O}(N^2\cdot |\set{F}|)$, $N$: population size). A quantitative perspective of such performance gap is shown in Fig.~\ref{Fig:PARETO_FRONT_MOEA_MDA_GAINS}. The blue/circle pattern corresponds to the gain in terms of $f_2$ for each value of $f_1$ indicated in the left vertical axis as `Average capacity gain'. As a result of the combinatorial nature of NSGA-II, the gains are higher when network topologies are composed of less BSs, i.e., small values of $f_1$. The red/square pattern shows the capacity gain per cell, indicated in the right vertical axis. It can be seen that the gain of using MOEA is around $1\,\text{Mbps/cell}$ in topologies with less than 20 active BSs~($f_1\leq20$). Hence, the use of MOEAs implies better network topologies in cases where the computational complexity can be afforded. The resulting Pareto Front by solving~(\ref{OP:CSO}), for ($i=1,\,j=3$) and ($i=1,\,j=4$) in~(\ref{OP:CSO:MAIN}), are shown in Figs.~\ref{Fig:PARETO_FRONT_P5_NEL}~and~\ref{Fig:PARETO_FRONT_UPC_NEL}, respectively. The first case illustrates the impact of CSO on cell edge performance. Note that while Fig.~\ref{Fig:PARETO_FRONT_MOEA_MDA} shows a fairly linear growth of the average network capacity with the number of active cells, Fig.~\ref{Fig:PARETO_FRONT_P5_NEL} indicates that cell edge performance (represented by $f_3$) is substantially improved only by network topologies featuring a higher number of active cells ($f_1\geq27$). This result clearly suggests that mechanisms for Intercell Interference Coordination~(ICIC) should be applied together with CSO in cases of low load conditions to improve the QoS of cell edge users. Fig.~\ref{Fig:PARETO_FRONT_UPC_NEL} illustrates the impact of CSO on the power consumption of users (uplink). As it was mentioned, the goal is not to determine exact uplink power consumption figures, but to create means for comparison among network topologies with different number of active BSs. Thus, a normalized version of $f_4$ (see~\ref{Eq:CSO_fUPLPC}) is considered. As it can be seen, it turns out that the relationship between the number of active BSs and the resulting uplink (open-loop-based) power consumption is highly nonlinear, being the energy expenditure considerably high in sparse network topologies~($f_1<15$). Hence, in scenarios where the lifetime of devices should be maximized (sensor networks), the use of CSO is not clear. Recall that uplink link budget is also considered as a coverage criterion.

		To close this subsection, Fig.~\ref{Fig:CELL_COUPLING_BASED_ANALYSIS} shows the results corresponding to the solution of~(\ref{OP:CSO}) for the objective functions introduced in Section~\ref{Sec:PropFramework_ProbForm_FL}~($f_5\,\text{and}\,f_6$). According to Definition~\ref{Def:NetCap}, and given the spatial demand distribution~$\myvec{\Gamma}$~(see Fig.~\ref{Fig:STD}), \mbox{$\mathbb{E}\{\lambda\}=115.0\,\text{ms}$} and \mbox{$\mathbb{E}\{\mu\}=119.2\,\text{s}$} yield a demand volume ($V$) equal to $V_{\text{Cap}}$. The resulting load sharing patterns (obtained by means of Algorithm~\ref{Alg:IterAproxLoads}) for \mbox{$V=V_{\text{Cap}}$} and \mbox{$V=0.5\cdot V_{\text{Cap}}$} are shown in Fig.~\ref{Fig:CV_LOADS}. Note that increasing~$V$ results in higher load dispersion. To quantify this, Fig.~\ref{Fig:CV_PC_LOAD} shows the impact of~$V$ on the Coefficient of Variation~(CV) of the loads~($f_6$). The associated load-dependent power consumption~($f_5$) is also indicated. Note that $f_5$ and $f_6$ are maximized when \mbox{$V=V_{\text{Sat}}$} and \mbox{$V=V_{\text{Cap}}$}, respectively. As expected, the load dependent power consumption ($f_5$) is maximized when $\bar{\boldsymbol{\alpha}}\geq\myvec{1}$, i.e., $V\geq V_{\text{Sat}}$. The dependence of $f_6$ on $V$ is explained by the strong nonlinearity of~(\ref{Eq:NELoadStatistically_2}) and the fact that, from the load-coupling point of view, $\bar{\boldsymbol{\alpha}}\leq\myvec{1}$, and hence, no change is expected after \mbox{$V=V_{\text{Sat}}$}. The results shown in Figs.~\ref{Fig:CV_LOADS}~and~\ref{Fig:CV_PC_LOAD} are obtained for $\myvec{x}=\myvec{1}$, i.e., when all the BSs are active. The joint optimization of $f_5$~and~$f_6$ is shown in Fig.~\ref{Fig:CV_PC_MO}. As it can be seen, there is a conflicting relationship between them. The attributes of the \textit{extreme} solutions ($\myvec{x}_{\text{A}}$ and $\myvec{x}_{\text{B}}$) in the Pareto Front are indicated. There is also a certain correlation between the objectives ($f_5$ and $f_6$) and the number of active cells~(NAC). The topology with the lowest energy consumption~($f_5$) requires less active BSs but it has the highest load dispersion ($f_6$). Note the difference between the highest and lowest loaded BS in $\myvec{x}_{\text{A}}$. In contrast, the best load balancing~($\myvec{x}_{\text{B}}$) involves more active BSs, and hence, worst values of $f_5$. A comparison among solutions obtained through each ICI model, FL and LC, is provided~next.

\subsection{System level simulations}\label{Sec:NumResults_SysLevSim}
As indicated earlier, solving (\ref{OP:CSO}) results in a set of Pareto efficient (nondominated) network topologies  that are specific for either a spatial service demand distribution ($\set{X}_{\text{FL}}$: full-load) or a service demand conditions, i.e., spatial demand distribution plus volume ($\set{X}_{\text{LC}}$: load-coupling). Recall that $\set{X}_{\text{FL}}$ is obtained by joint optimizing $f_1$ and $f_2$ in (\ref{OP:CSO}) for a given spatial demand distribution ($\myvec{\Gamma}$), while obtaining $\set{X}_{\text{LC}}$ involves the joint optimization of $f_5$ and $f_6$ in (\ref{OP:CSO}) for a given $\myvec{\Gamma}$ and $V$ (volume). Note that, the `full-load' analysis is volume-independent, and hence, it does not require specify $V$ (full load is assumed for the active cells).  
			\begin{figure*}[t]
	    		\centering
	    		\subfloat[Power consumption.]
	    		{\label{Fig:COMPARATIVE_C_NC_LOAD}
	    		\includegraphics[width = 0.245\textwidth]{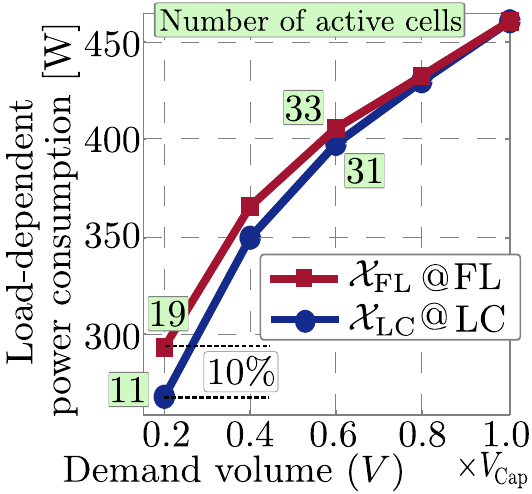}}
					\subfloat[$V=0.2\cdot V_{\text{Cap}}$.]
	    		{\label{Fig:COMPARATIVE_C_NC_LOAD_DOT2}
	    		\includegraphics[width = 0.364\textwidth]{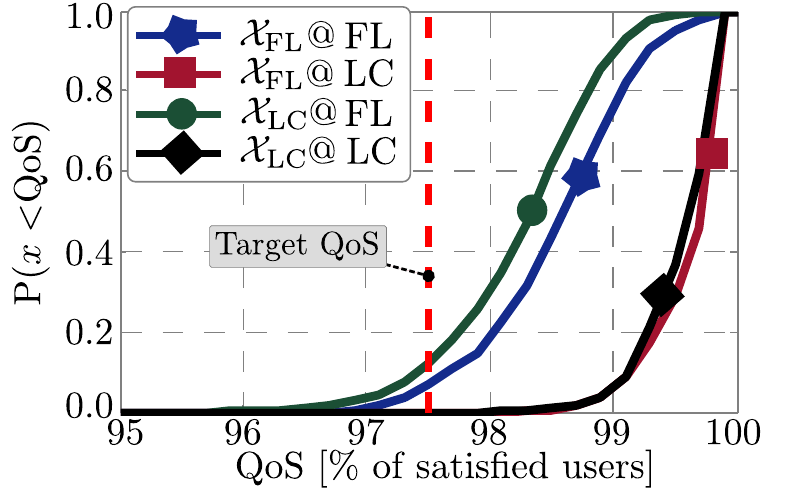}}
					\subfloat[$V=0.6\cdot V_{\text{Cap}}$.]
	    		{\label{Fig:COMPARATIVE_C_NC_LOAD_DOT6}
	    		\includegraphics[width = 0.364\textwidth]{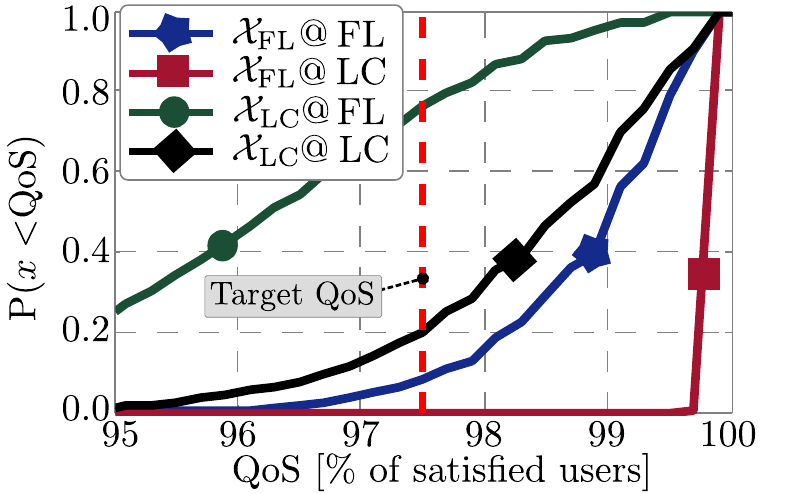}}
	    		\vspace{-0.05cm}\caption{Comparative analysis: Load-coupling vs. Full ICIC MO optimization.}
	    		\label{Fig:COMPARATIVE}    		
			\end{figure*}	
							\begin{figure*}[t]
	    		\centering
	    		\subfloat[Number of active cells.]
	    		{\label{Fig:BENCH_NEL}
	    		\includegraphics[width = 0.475\textwidth]{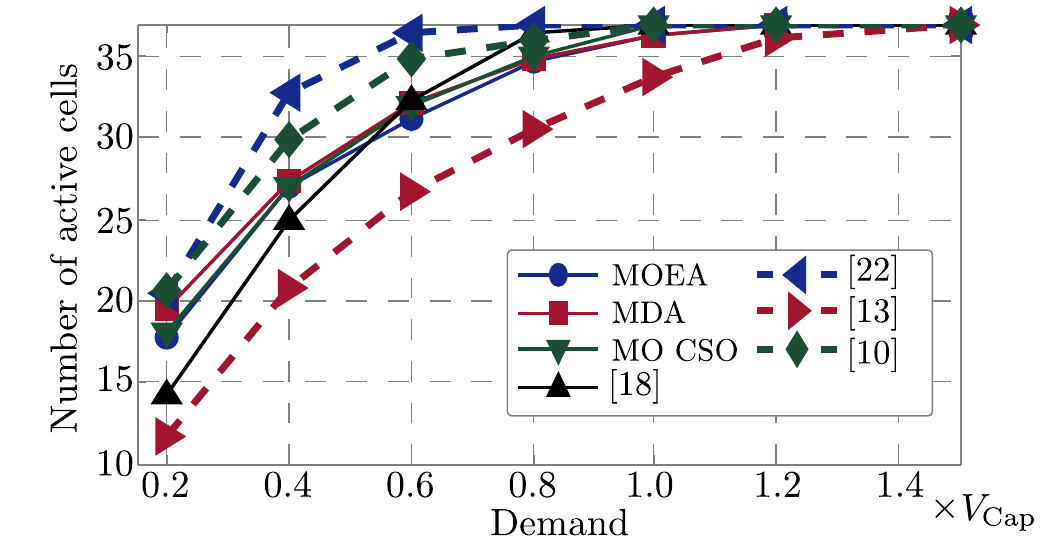}}
					\subfloat[QoS.]
	    		{\label{Fig:BENCH_QoS}
	    		\includegraphics[width = 0.475\textwidth]{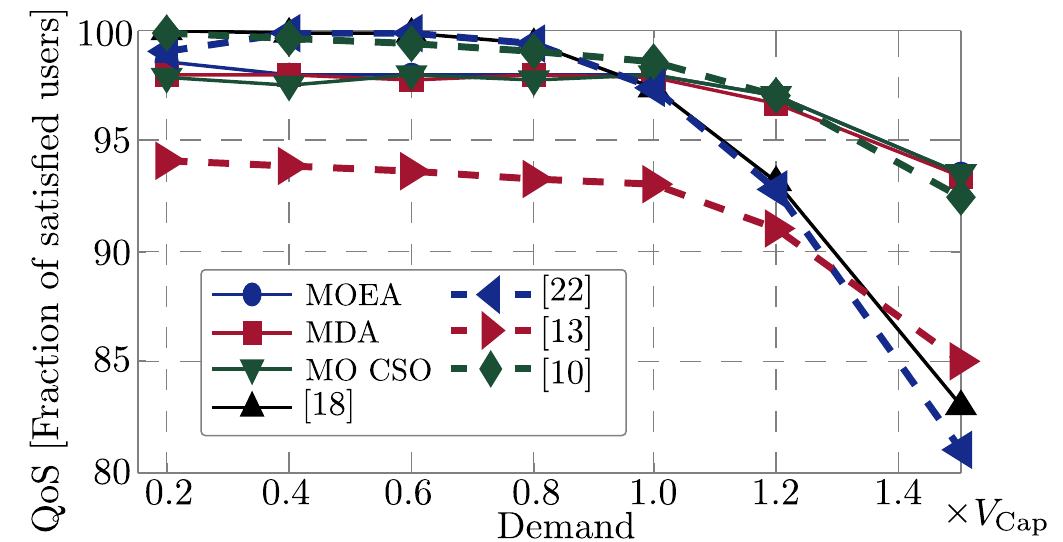}}\\[-0.05cm]
					\subfloat[Key performance indicators.]
	    		{\label{Fig:BENCH_BARS}
	    		\includegraphics[width = 0.780\textwidth]{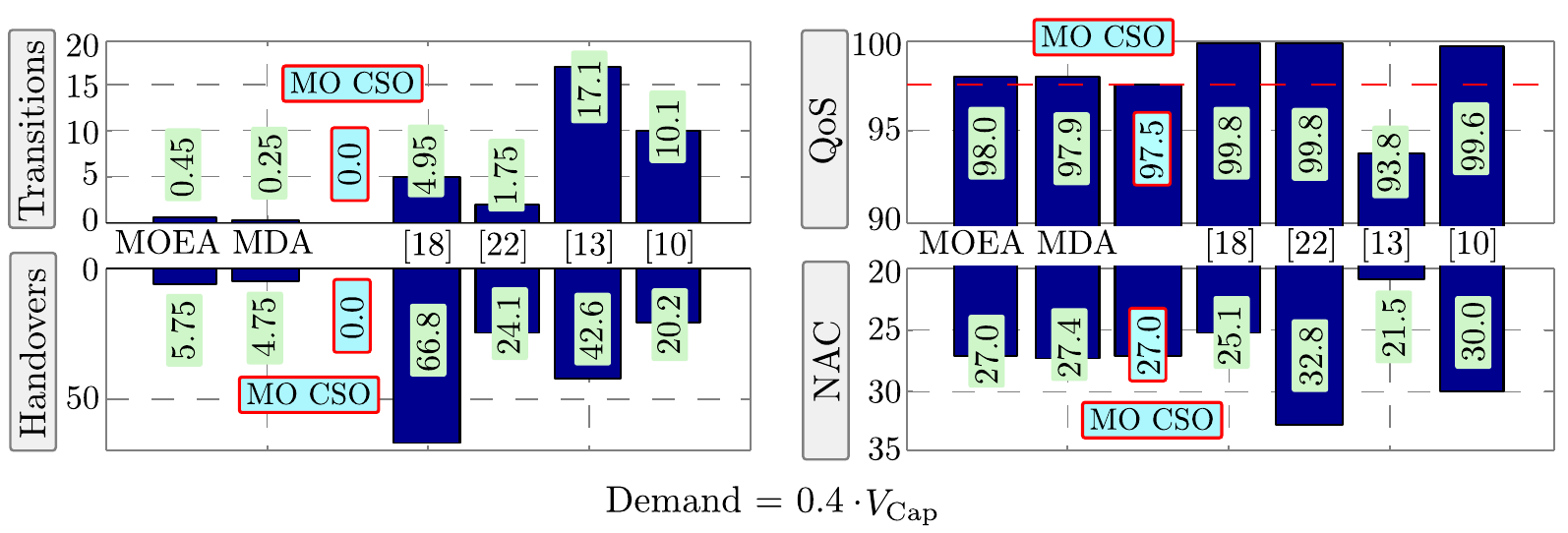}}
	    		\vspace{-0.1cm}\caption{Performance comparisons.}
	    		\label{Fig:PER_COMP}    		
			\end{figure*}	
Thus, in order to evaluate these solutions by means of system level simulations, it is initially assumed that at each QoS checking interval (evaluation parameters are shown in Table~\ref{TableEvaluationSetting}), the (nondominated) network topologies of each set ($\set{X}_{\text{FL}}$~and~$\set{X}_{\text{LC}}$) are all applied and evaluated. The goal is to create QoS statistics for each network topology and load condition. Then, the network topology that is able to provide the desired QoS ($Q\,\%$ of users are satisfied $Q\,\%$ of time) is selected and applied (as indicated in Subsection~\ref{Sec:PropFramework_ConceptualDesign}). 
The comparative assessment is shown in Fig.~\ref{Fig:COMPARATIVE}, where the legends indicate the set the applied network topology belongs to ($\set{X}_{\text{FL}}$ or $\set{X}_{\text{LC}}$) and the ICI model (FL or LC) used in the system level trials. Fig.~\ref{Fig:COMPARATIVE_C_NC_LOAD} shows the load-dependent power consumption of each network topology. Clearly, from the CSO point of view, the topologies in $\set{X}_{\text{LC}}$ result in lower power consumption as they feature less active BSs (NAC is indicated in green boxes) given that the load-coupling model predicts better SINR than full-load~(see~Fig.~\ref{Fig:SINR_Bound}), and hence, network capacity is favored. However, as $V$ increases, both models become somehow equivalent as the loads tend to 1; as a result, the energy consumption is quite similar. Figs.~\ref{Fig:COMPARATIVE_C_NC_LOAD_DOT2}~and~\ref{Fig:COMPARATIVE_C_NC_LOAD_DOT6} show the QoS level (in terms of the number of satisfied users) that is obtained with the selected solution of each set for $V=0.2\cdot V_{\text{Cap}}$ and $V=0.6\cdot V_{\text{Cap}}$, respectively. The results make evident that the performance of the network topologies in $\set{X}_{\text{LC}}$ is severely degraded if the ICI levels become higher than the ones from which they were calculated for, see~$\set{X}_{\text{LC}}$@FL (full ICI). Indeed, the performance of these solutions is sensitive to variations from the mean values (that happens when considering snapshots) in moderate-to-high load conditions, even when the load-coupling based ICI is considered, as seen in Fig.~\ref{Fig:COMPARATIVE_C_NC_LOAD_DOT6} for $\set{X}_{\text{LC}}$@LC. On the other hand, the network topologies in $\set{X}_{\text{FL}}$ provide consistent performance when they are evaluated under full load ($\set{X}_{\text{FL}}$@FL), and obviously, provide an even better performance under load-coupling~($\set{X}_{\text{FL}}$@LC) for both demand volumes. Hence, given that the energy consumption gain is in the order of $10\,\%$ in the best case, it can be concluded that the full load model provides a competitive and somehow \textit{safer} energy-saving vs. QoS tradeoff in the context of CSO. The proposed CSO scheme can use either approach. Summarizing:
{\renewcommand{\labelitemi}{$\checkmark$}	
					\begin{itemize}
	\item The MO for FL, i.e., $f_1$ and $f_2$ in (\ref{OP:CSO}), is volume-independent; offline system level simulations are required for each load condition ($V$), and energy saving is smaller in comparison to~LC.
	\item In MO for LC, i.e., $f_5$ and $f_6$ in (\ref{OP:CSO}), is volume dependent; different offline optimization procedures are required for each load condition ($V$), and energy saving is larger in comparison to~FC.
\end{itemize}
\subsection{Performance comparison}\label{Sec:NumResults_PerComp}
	In order to provide a wide perspective of the merit of the CSO framework presented herein, several recent/representative CSO schemes have been used as baselines. Obviously, an exhaustive comparison is not feasible. However, the idea is to illustrate some \textsl{pros} and \textsl{cons} of different approaches and the impact of some design assumptions. The following benchmarks are considered:
					\begin{itemize}[leftmargin=0.3cm]
					\item \textsl{Cell zooming}:  It was proposed in~\cite{05:00125}. The idea is to sequentially switch-off BS starting from the lowest loaded. The algorithm ends when a cell cannot be switched-off because at least one user cannot be \mbox{re-allocated.}
					\item \textsl{Improved cell zooming}: This scheme is presented in~\cite{04:00212} and it is similar to the one in~\cite{05:00125}, but it includes a more flexible termination criterion that allows to check more cells before terminating, and so, more energy-efficient topologies can be found.
					\item \textsl{Load-and-interference aware CSO}: The design of this CSO scheme presented in~\cite{04:00341} takes into account both the received interference and load of each cell to create a ranking that is used to sequentially switch-off the cells whose load is below a certain threshold. 
						\item \textsl{Set cover based CSO}: The CSO scheme proposed in~\cite{04:00342} relies on the idea of switch-on BS sequentially according to a certain sorting criterion. In this work, the sorting criterion is based on the number of users a cell can served in the SNR regime. 
						\end{itemize}  
The performance comparison is shown in Fig.~\ref{Fig:PER_COMP}. To make the comparison fair, the full-load ICI conditions are considered. Figs.~\ref{Fig:BENCH_NEL}~and~\ref{Fig:BENCH_QoS} show the average number of active cells and QoS (for different service demand volumes), respectively. As it can be seen, the best energy saving is obtained by~\cite{04:00341}, although at the expense of QoS degradations. This is due to the fact that in~\cite{04:00341}, users can be easily put in outage. In contrast, CSO schemes such as~\cite{04:00212, 05:00125, 04:00342} provide the desired QoS (as long as $V\leq V_{\text{Cap}}$) since CSO decisions require associating all users. However, this results in an increment in the average number of active cells with respect to~\cite{04:00341}. The schemes labeled as `MDA' and `MOEA' correspond to the (infeasible) dynamic selection of network topologies from the sets $\set{X}_{\text{FL}}$ obtained through Algorithms~\ref{Alg:CSO:MinDist} and NSGA-II, respectively, which are shown as reference. The performance of the proposed MO CSO is indicated by red boxes and labeled as `MO~CSO'. As it can be seen, the proposed scheme provides an excellent tradeoff between the required number of active cells and the obtained QoS, especially when $V\geq V_{\text{Cap}}$ where the performance (QoS) of other CSO is compromised. However, the most significant enhancement in the proposed scheme is its feasibility. Fig.~\ref{Fig:BENCH_BARS} shows four performance indicators: transitions, handovers, QoS, and NAC. Given that the network topologies are calculated offline, they can be evaluated extensively by means of system level simulations (under a wide range of coverage criteria and conditions) to further guarantee their real-time performance, i.e., the operator can select topologies with more active cells rather than the ones which strictly need to guarantee QoS. Therefore, the selected network topologies can be applied (without real-time complexity) during periods of time in which service demand is described by $\myvec{\Gamma}$; as a result, no transitions or handovers are induced due to CSO. Hence, feasible yet effective CSO performance is achieved. As it was shown earlier, the proposed framework is generic, flexible, and no assumption are made in regards to, for instance, the cellular layout or objective functions; as a result, the framework is also suitable for small-cell deployments where irregular topologies and heterogeneous demand conditions are expected.

\subsection{Complexity analysis and calibration aspects}\label{Sec:NumResults_Complexity analysis}
To close this section, a complexity overview of the optimization algorithms is provided. 
According to~\cite{05:00087}, the complexity of NSGA-II is $\mathcal{O}(M\cdot N^2)$, where $M$ and $N$ correspond
 to the population size and the number of objective functions, respectively. In our case, $N=2$ and $M$ can be 
set depending on the scale of the problem. However, there is a consensus about the size of the population 
when using genetic algorithms, such as NSGA-II, and it is considered that during calibration populations
of 20 up to 100 individuals can be used. Values greater than 100 hardly achieve significant gains and
the same global convergence is obtained~\cite{spall2005introduction}. Regarding Algorithm~\ref{Alg:CSO:MinDist},
it's complexity is $\mathcal{O}(L^2)$, where $L$ is the number of cells in the network. In practice, $L^2\ll M\cdot N^2$ which is a significant reduction 
in terms of complexity that comes at expense of some performance. In evolutionary algorithms, a termination criterion is usually defined/need. One metric 
used to measure the level of convergence is the the \textsl{hypervolume} indicator~\cite{08:00052}. It reflects the size of volume dominated
by the estimated Pareto Front. In this work, the seacrh is terminated if the improvement in the hypervolume is smaller than a threshold~(0.001\%) after a certain number of
generations (in this study, 20). Finally, crossover and mutation probabilities are set to $1$ and $1/L$ (one mutation per solution, on average), respectively, as indicated in Table~\ref{TableEvaluationSetting}.
}

\section{Conclusions and Research Directions}\label{Sec:Conclusions}			
CSO is a promising strategy that allows significant energy saving in cellular networks where both radio access network~(capacity supply) and service demand are heterogeneous. In this article~1)~CSO has been carefully analyzed considering coverage criteria, ICI models, and practical aspects, such as network-initiated handovers and on-off/off-on transitions, and~2)~a novel MO-based~CSO scheme has been introduced. The proposed solution succeeds in minimizing the number of transitions and handovers caused by the CSO operation and it is able to operate without need for heavy computational burden as the core processing is done offline. In addition, a cluster based-operation have been proposed to allow for semi-distributed implementation. The results show that, when compared with previous proposals, the proposed solution provides competitive performance in terms of QoS and energy saving while offering clear advantages from the feasibility perspective as it reduces the number of handovers and transitions. 
The results also highlight the importance of considering coverage criteria (in downlink and uplink) and pay attention to the selection of operational parameters, e.g., the power allocated to PS (typically used as criterion for coverage). \vspace{-0.05cm}

A comparative analysis between ICI models (full-load and load-coupling) indicates that the full-load assumption is a \textit{safe} approach in the context of CSO as it provides natural protection against deviations from average load values that are 1) used as input of the algorithm, and~2)~inherent of real time operation, i.e., discrete realizations of users. 
The impact of CSO on the power consumption of UE has also been studied. The results indicate that sparse topologies (few active BSs) have a significant impact on uplink power consumption, and hence, CSO is not suitable for scenarios with energy-sensitive devices such as sensor networks. 

 Research on topology adaptation has still a long way until its maturity. Feasible and effective techniques for traffic pattern recognition to complement CSO are still in infancy. It is our strong belief that CSO, as a promising approach to \textit{greener} networks, is a key piece of a more general set of capabilities that will appear in 5G networks, also including promizing and disruptive concepts, such as Downlink Uplink Decoupling~(DUDe). DUDe, where user equipment can transmit and receive to and from different base stations, is indeed, a clear research direction from the perspective of CSO, where both uplink and downlink could be considered as \textit{independent} networks.  

  \section*{Acknowledgment}
The authors would like to thank Tamer Beitelmal from Carleton University, and Dr. Ngoc Dao from Huawei Canada Research Centre, for their valuable feedback.	This work was supported by the Academy of Finland (grants 287249 and 284811).	Mario García-Lozano is funded by the Spanish National Science Council through the project \mbox{TEC2014-60258-C2-2-R}.

\vspace{-0.4cm}
\appendices
			\section{Iterative approximation of cell loads}\label{App:LoadCoupling}
\begin{figure}[t]
	    		\centering
	    		\subfloat[Bounds of SINR distributions.]
	    		{\label{Fig:SINR_Bound}
	    		\includegraphics[width = 0.43\textwidth]{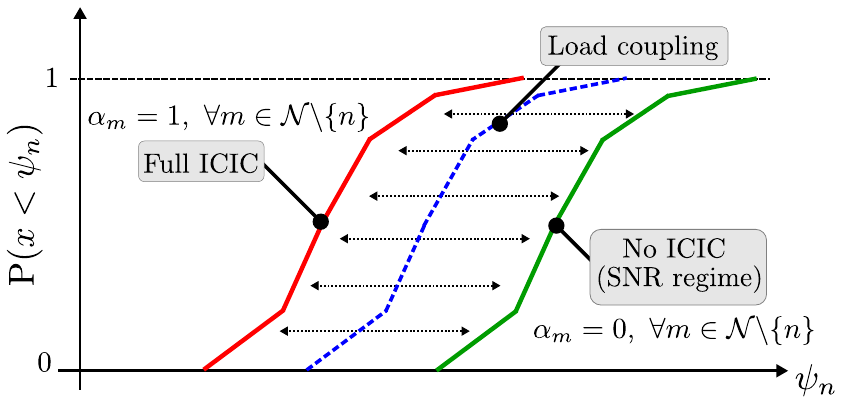}}\\[-0.015cm]
					\subfloat[Iterative load estimation.]
	    		{\label{Fig:ITER_LOAD}
	    		\includegraphics[width = 0.43\textwidth]{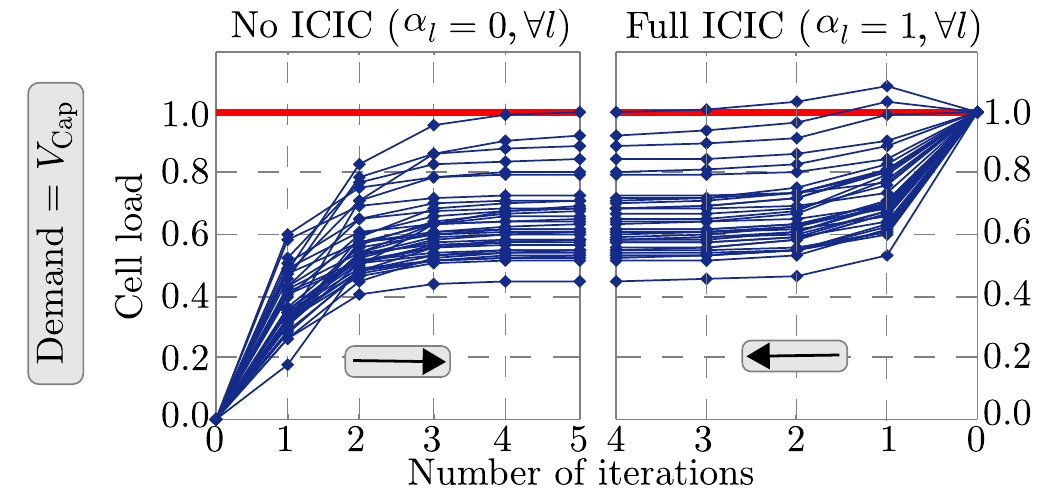}}
	    		\vspace{-0.10cm}\caption{Impact of cell load coupling on SINR distributions.}
	    		\label{Fig:ITER_LOAD_ALG}    		
			\end{figure}

In order to estimate the average load vector ($\bar{\boldsymbol{\alpha}}$), Algorithm~\ref{Alg:IterAproxLoads} is proposed.
		Basically, the estimation of the average load at each cell ($\bar{\alpha}_l$) is refined through each iteration comprising Lines~\ref{Alg1_l4}~to~\ref{Alg1_OuterLoop2}.  In line~\ref{Alg1_Load}, the function~\ttfamily Load()\rmfamily~estimates each $\bar{\alpha}$, based on (\ref{Eq:NELoadStatistically_2}), from the values of  previous iterations (where \mbox{$\bar{\alpha}_l=\text{min}\{1,\bar{\alpha}_l\}$}) and the ones that have been just updated in the current iteration (this \textit{fast} update is done in line~\ref{Alg1_innerLoop2}). Fig.~\ref{Fig:ITER_LOAD_ALG} illustrates the motivation and performance of Algorithm~\ref{Alg:IterAproxLoads}. Basically, the use of load coupling provides a more accurate estimation of ICI levels in the network as shown in Fig.~\ref{Fig:SINR_Bound}. Note that the use of Full ICIC represents a more \textit{conservative} approach. Fig.~\ref{Fig:ITER_LOAD} shows that Algorithm~\ref{Alg:IterAproxLoads} only requires few iterations to converge and that this depends on the starting point, but in any case convergence is fast.
			\begin{algorithm}[t]
 \SetAlgoLined
								\SetKwInOut{Input}{Inputs}
								\SetKwInOut{Output}{Output}	
								\SetKwFunction{Load}{Load}	
								\SetKwFunction{Sort}{Sort}
								\BlankLine\small
							\Input{{\small~All relevant information and $\epsilon$ (termination).\\
										}} 
							\Output{{\small~$\bar{\boldsymbol{\alpha}}\in\spc{R}_{+}^{|\set{L}|}$: Load vector.
										}}
\BlankLine
		$\boldsymbol{\bar{\alpha}}^{0}\leftarrow\myvec{1}$; $\boldsymbol{\epsilon}\leftarrow\epsilon\cdot\myvec{1}$; $k\leftarrow 0$\tcc*[r]{{\footnotesize Initializing}}\label{Alg1_l1}\vspace{-0.0cm}
\Repeat{$\boldsymbol{\epsilon}>\frac{|\boldsymbol{\bar{\alpha}}^{k}-\boldsymbol{\bar{\alpha}}^{*}|}{\boldsymbol{\bar{\alpha}}^{*}}$}{\vspace{-0.0cm}
			$k\leftarrow k+1$\label{Alg1_l4}\;\vspace{-0.0cm}
			$\boldsymbol{\bar{\alpha}}^{*}\leftarrow \boldsymbol{\bar{\alpha}}^{k-1}$\label{Alg1_l5}\;\vspace{-0.0cm}
			\For{$l=1:|\set{L}|$}
				{\vspace{-0.0cm}
					$\bar{\alpha}_l^{k}\leftarrow$\Load{~$\{\bar{\alpha}_{j}^{k-1}:(\forall j\in\set{L}) \wedge (j\neq l)\}$}\label{Alg1_Load}\;\vspace{-0.0cm}
					$\bar{\alpha}_l^{k-1}\leftarrow\bar{\alpha}_l^{k}$\tcc*[r]{{\footnotesize Fast update}}	\label{Alg1_innerLoop2}\vspace{-0.0cm}				
				}	\vspace{-0.0cm}
			$\boldsymbol{\bar{\alpha}}^{k}\leftarrow[\bar{\alpha}_0^{k}\,\cdots\,\bar{\alpha}_{L-1}^{k}]$\tcc*[r]{{\footnotesize Update: iteration $k$}}\label{Alg1_OuterLoop2}	\vspace{-0.0cm}		
}\vspace{-0.0cm}
\textbf{return}~~$\boldsymbol{\bar{\alpha}}^{k}$\tcc*[r]{{\footnotesize Return estimated load vector}}	\vspace{-0.0cm}
		\caption{Iterative approximation of cells load.}
		\label{Alg:IterAproxLoads}
		\end{algorithm}

\end{document}